\documentclass[10pt,letterpaper]{article}
\usepackage[top=0.85in,left=2.75in,footskip=0.75in]{geometry}

\usepackage{CJKutf8} 
\usepackage{pdfpages} 

\usepackage{amsmath,amssymb}

\usepackage{changepage}

\usepackage{textcomp,marvosym}

\usepackage{cite}

\usepackage{nameref,hyperref}

\usepackage[right]{lineno}

\usepackage{microtype}
\DisableLigatures[f]{encoding = *, family = * }

\usepackage{array}

\newcolumntype{+}{!{\vrule width 2pt}}

\newlength\savedwidth

\raggedright
\usepackage[aboveskip=1pt,labelfont=bf,labelsep=period,justification=raggedright,singlelinecheck=off]{caption}

\makeatletter
\renewcommand{\@biblabel}[1]{\quad#1.}
\makeatother

\usepackage{lastpage,fancyhdr,graphicx}
\usepackage{epstopdf}
\fancyheadoffset[L]{2.25in}
\fancyfootoffset[L]{2.25in}
\begin{document}
\vspace*{0.2in}

\begin{flushleft}
{\Large
\textbf\newline{Competition for popularity and  interventions on a Chinese microblogging site} 
}
\newline
\\
Hao Cui\textsuperscript{1}
János Kertész\textsuperscript{1*}
\\
\bigskip
\textbf{1} Department of Network and Data Science, Central European University, Quellenstrasse 51, A-1100, Vienna, Austria
\\
\bigskip

%
%





* kerteszj@ceu.edu

\end{flushleft}
\section*{Abstract}

Microblogging sites are important vehicles for the users to obtain information and shape public opinion thus they are arenas of continuous competition for popularity. Most popular topics are usually indicated on ranking lists. In this study, we investigate the public attention dynamics through the Hot Search List (HSL) of the Chinese microblog Sina Weibo, where trending hashtags are ranked based on a multi-dimensional search volume index. We characterize the rank dynamics by the time spent by hashtags on the list, the time of the day they appear there, the rank diversity, and by the ranking trajectories.  We show how the circadian rhythm affects the popularity of hashtags, and observe categories of their rank trajectories by a machine learning \textcolor{black}{clustering} algorithm. By analyzing patterns of ranking dynamics \textcolor{black}{using various measures,} we identify anomalies that are likely to result from the platform provider's intervention into the ranking, including the anchoring of hashtags to certain ranks on the HSL. We propose a simple model of ranking that explains the mechanism of this anchoring effect. 
\textcolor{black}{We found an over-representation of hashtags related to international politics at 3 out of 4 anchoring ranks on the HSL, indicating possible manipulations of public opinion.} 


\section*{Introduction}

Studying public attention is important from various aspects including governance, public security, marketing, and pandemic management~\cite{aksoy2020public, dyer2020public}. With the development of digital technology, social media have permeated into the society and become an inevitable source of information for many. An increasing part of the public obtains the latest information from social media, expresses opinions and attitudes there. This results in a strong competition for visibility and popularity partly because of their obvious relation to power and financial gain. Due to the large number of users and the volume of activity, changes in popularity happen at a high pace. As the novelty of the hot topics fades rapidly with time~\cite{wu2007novelty}, public attention shifts to new, current trending topics. Popularity trends on social media can be regarded as a proxy of collective public attention. 
In the social media ecosystem, microblogging sites are important platforms because of the conciseness of the messages, the high turnover speed and the open online social networks of their users. Microblogging sites are complex interacting systems where users generate content and disseminate information through posts, reposts, comments, likes\textcolor{black}{,} and mentions, and where emergent phenomena occur like macroscopic collective behaviors triggered by some pieces of information \cite{eom2015twitter}. 

Detecting and indicating the trends in popularity is important since they reflect \textcolor{black}{the} most concerned issues by the public in real time, which is highly relevant for a number of issues, including the popularity of cultural products, market changes \cite{annamoradnejad2019comprehensive}, government policy-making, and elections~\cite{McGregor2017politics}. 
In order to inform customers, microblogging site providers like Twitter or Sina Weibo present trending lists based on some statistical measures. The lists have twofold roles: First, they indicate popularity of given topics, second, they boost popularity of those topics which manage to get to the list. Therefore there is a competition in getting to the list and staying there as long as possible. 

Twitter trending list shows hot topics around the globe. Research studies have analyzed the dynamics of trending topics through comprehensive statistical analysis from the aspects of lexical composition, trending time, trend re-occurrence \cite{annamoradnejad2019comprehensive}, etc. There have been different factors identified, which contribute to the success of a topic, like novelty of the piece of information and the resonance level of the messages spread as well as the influence of certain members of the propagating network~\cite{romero2010influence}. The evolution of Twitter trends is characterized by phases of burst, peak and fade~\cite{asur2011trends} and the patterns of temporal evolution of popularity of hashtags have been ordered into six different categories~\cite{yang2011patterns}.

The economic and political relevance of popularity of items on online media is an incentive to the service providers to intervene into the trending lists. A linear influence model~\cite{yang2010modeling} was introduced to capture the network effect on endogenous diffusion of hashtags on Twitter trending list and demonstrate evidences of manipulation~\cite{zhang2016twitter} on the observed dynamics. Certain trending topics on Twitter may be opportunistically targeted with desirable qualities by spammers~\cite{stafford2013evaluation}. 
Recent studies on Twitter trends have found likely presence of coordinated campaigns in AstroTurf version to influence and manipulate public opinion during the COVID-19 crisis in Mexico~\cite{pina2022coordinated}. 

The Chinese microblogging site Sina Weibo is geographically more limited but larger than Twitter in terms of number of daily and monthly active users~\cite{mau}. Although it has generated less academic publications than Twitter, it has attracted considerable research attention due to its enormous user participation and profound role in mainland China where Twitter is blocked~\cite{bamman2012censorship}. The Hot Search List (HSL) on Sina Weibo, with a role similar to Twitter trending list, is a major source for people from mainland China to obtain real-time information about the popularity of topics. Research on Weibo hot topics has focused on topic dynamics from the perspective of time, geography, demographics, emotion, retweeting, and correlation~\cite{fan2015topic}, on similarities and differences to Twitter \cite{yu2015trend, yu2011trends}, emergence mechanisms~\cite{cui2022born}, patterns of popularity evolution~\cite{kong2018exploring}, prediction~\cite{zhao2014short, liu2017interdisciplinary, zhou2017predicting}, social emotions and diffusion patterns~\cite{fan2014anger} as well as impact of censorship~\cite{chen2013tweeting}.

The ranking of trending contents on social media changes over time, following the rise and fall of public attention dynamics: Old trends vanish and new trends emerge. The search volume\textcolor{black}{s} for hashtags indicate their popularity and on Sina Weibo this quantity is \textcolor{black}{supposed to be one of} the main underlyings for the ranking list HSL, as reflected in the name of the list. 

The exposure of hashtags on the HSL has a great promotional effect thus many are keen to be on the list, resulting in strong competition and manipulation attempts. Studies reveal that hashtags from different topical categories differ in time length of prehistory (from birth till first appearance on the HSL) and the types of accounts involved in the propagation~\cite{cui2022born}. Celebrity and entertainment related hashtags are often associated with marketing accounts~\cite{zhang2015study} which can be influenced by social capital~\cite{cui2022born}. Recent findings indicate possibility of algorithmic intervention~\cite{cui2021attention} from the platform provider towards COVID-related hashtags on the HSL during the COVID-19 pandemic. Research indicated that human editorial decisions were involved in the curation of Weibo trending topics with the aim of increasing user engagement~\cite{Li2021Intervention} and that Weibo actively facilitates the production and spread of online contention to attract more users through a range of recommendation mechanisms built into the platform, including the trending topic list and channels such as Sina-owned official accounts~\cite{Li2021Intervention}. As contents on social media can influence social perception~\cite{perloff2009mass}, studying the dynamics of both ranking position and duration of the hashtags on the HSL can deepen our understanding about the dynamics of public attention, its relation to hashtag prehistory, and reveal ways of interventions on social media platforms.  

In this paper, we study the attention dynamics on Sina Weibo by digging deeper into the characteristics of the hashtags on the HSL. 
We describe the patterns of the ranking position and duration of hashtags, introduce ranking trajectory classification, and uncover its relationship with the prehistory and the time of the day when the hashtag first appears on the HSL. 
We identify anomalies which can be attributed to intervention into the ranking by the service provider and propose a model of anchoring to explain the anomalous ranking dynamics of hashtags on the HSL. \textcolor{black}{We categorize the anchored hashtags based on their semantic meaning to uncover possible motivation for interference by the service provider.} 

\section*{Materials and methods}
\subsection*{Sina Weibo Data}

Sina Weibo Hot Search List is a convenient tool for netizens to follow hot topics, news events and celebrity gossip, and has gone through several reforms. In 2014, a real-time Hot Search List was launched on the client with an updating frequency of once every ten minutes~\cite{hslrule}, allowing users to see the latest hot information anytime, anywhere. The updating frequency later accelerated to every minute \cite{hslrule} in 2017. Until 2021\textcolor{black}{,} there used to be one or two advertisement rank positions included in the top 50 ranks{\footnote{Since 2021, Weibo went through another reform that the advertisements hashtags are randomly placed between ranks 3 and 4 or/and ranks 6 and 7, without a rank number associated in front of the advertisement, resulting in maximum 52 hashtags on the HSL, excluding the imposed positive energy recommendation position (hot search top) above all ranks which is often promotion for positive contents.  Sina Weibo has also separated a new list specifically for entertainment and celebrity related hashtags, though in the normal HSL there are still entertainment related hashtags. Our data was collected from the period before the reform.}}.
\textcolor{black}{Sina Weibo HSL claims to publish the ranking of the top 50 most popular hashtags based on a multi-factor index~\cite{Weiboindex}. Due to its opaque way of determining and ranking hashtags on the HSL, Weibo has been target of criticism for making financial gains as the HSL serves as an advertising tool to boost popularity. Sina Weibo responded to the criticism on 23 August 2021 and released what it called the rule of capturing the ``hotness" $H_i$ of a hashtag $i$ at a certain time~\cite{hslrule}, in the form of the following formula:
\begin{equation}\label{eq:1}
    H_i=(S_i+D_i+R_i)\times I_i,
\end{equation}
where search hotness $S_i$ refers to the search volume, including manual input search and click-and-jump search, discussion hotness $D_i$ is for the amount of discussion, including original posting and reposting, reading hotness $R_i$ is the spreading popularity equal to the number of readings, reflecting the spread of hotspots in the Weibo system,
and interaction hotness $I_i$ refers to the interaction rate of hot search results page for that hashtag.}

Sina Weibo HSL contains the names of the hashtags, their ranks and the search volume hotness which is the base of the ranking (see Eq.~(\ref{eq:1})).
We crawled the data from Sina Weibo HSL, with a frequency of $\Delta t=5$ minutes from 22 May 2020 to 29 September 2020. Since the commercial advertisements randomly occupied the HSL at the third and the sixth ranks, in order to get a constant length of non-advertisement hashtags on the HSL at each timestamp, we removed all the advertisement hashtags which are labeled with ``Recommendation (\begin{CJK*}{UTF8}{gbsn}荐\end{CJK*})", re-ranked the original HSL and took the top $L=48$ hashtags for each timestamp, with $L$ being the length of the list. 
Weibo was punished by the cyberspace authority of China to suspend the update of HSL for one week in June 2020 due to its interference with online
communication~\cite{suspension}, which causes a gap in the data (see Fig.~\ref{fig:fig1}).
We then did our major analysis based on the data after the punishment. We took all the hashtags that have appeared on the HSL in a two month period from 17 July to 17 September 2020, and crawled all the posts containing these hashtags in their prehistory from birth till first appearance on the HSL. \textcolor{black}{The datasets used in this research are available in a GitHub repository~\cite{cui2022github}.}

\subsection*{\textbf{Ranking dynamics}}
\subsubsection*{\textbf{Measures}}

A popular hashtag $i$ enters the HSL at time $t_i$ at \textcolor{black}{enter-}rank $r_i(t_i)$ \textcolor{black}{with $1\leq r_i(t_i) \leq L$} and disappears from it at time $T_i$ \textcolor{black}{at leave-rank $r_i(T_i)$. During the period $t_i\leq t \leq T_i$ the rank of this hashtag changes with time producing a trajectory $r_i(t)$ on the HSL.} 
In order to capture the ranking characteristics of hashtags at different ranks, we use the measure rank diversity~\cite{morales2016generic, iniguez2022dynamics}.
Rank diversity $D(k)$ measures the number of different hashtags at rank $1\leq k \leq L$ during a given period of time $t_{\rm{min}}\leq t \leq t_{\rm{max}}$: 
\begin{equation}
D(k)=\sum_t\sum_{i} \delta(k,r_i(t))\phi_{i,k}(t),
\end{equation} 
where $\delta(\cdot,\cdot)$ is the Kronecker delta and $\phi_i(t)$ is the indicator, which is $1$ if hashtag $i$ has not been at rank k until time $t$ and $0$ otherwise.

Rank diversity has been studied extensively. It is known that these quantities are characterized by profiles: For high ranks, their diversity have small values, while the behavior for lower ranks depends on whether the system is closed (only the rank changes but the items do not) or open (when items arrive on and leave from the list). In closed systems the dynamics at low ranks is also suppressed leading to low values of $D$ and a maximum at intermediate ranks, while in open systems these quantities grow monotonously, as it can be shown by simple diffusive models~\cite{morales2016generic, iniguez2022dynamics, morales2018rank}. An open system can be considered as a part of a very large closed system.

The duration $d_i$ of a popular hashtag on the HSL measures the time over which it is able to attract consistent public attention: 
\begin{equation}
    d_i = T_i - t_i.
\end{equation}
\textcolor{black}{Note, that ``high rank" means small rank value, i.e., the highest rank has rank 1.} 
The highest rank $r_i^{\rm{min}}$ of a hashtag measures its maximum relative ability to attract public attention during its whole lifetime on the HSL:
\begin{equation}
r_i^{\rm min} = \min_{t\in[t_i,T_i]}r_i(t).
\end{equation}

\subsubsection*{\textbf{Categorization of rank trajectories}}

The rank trajectory $r_i(t)$ is uniquely defined for $\forall$ hashtag $i$. Some hashtags have short lifetime on the HSL, others can attract popularity for a longer period of time; some go rapidly to high ranks, others never reach that level. Are there similarities between different shapes of the trajectories and can they be ordered into categories? Here we use machine learning techniques to find characteristic patterns in these rank trajectories. In order to deal with rank time series of different lengths, we use Dynamic Time Warping (DTW)~\cite{muller2007dynamic} as a similarity measure between two time series. DTW computes the best possible alignment between two time series. Then we use k-means clustering to find clusters of characteristic shapes. The computation was done using python tslearn package~\cite{JMLR:v21:20-091}. 

\section*{Results}
\subsection*{\textbf{Circadian patterns}}

\begin{figure}[htbp]
    \begin{center}
      \includegraphics[scale=0.098]{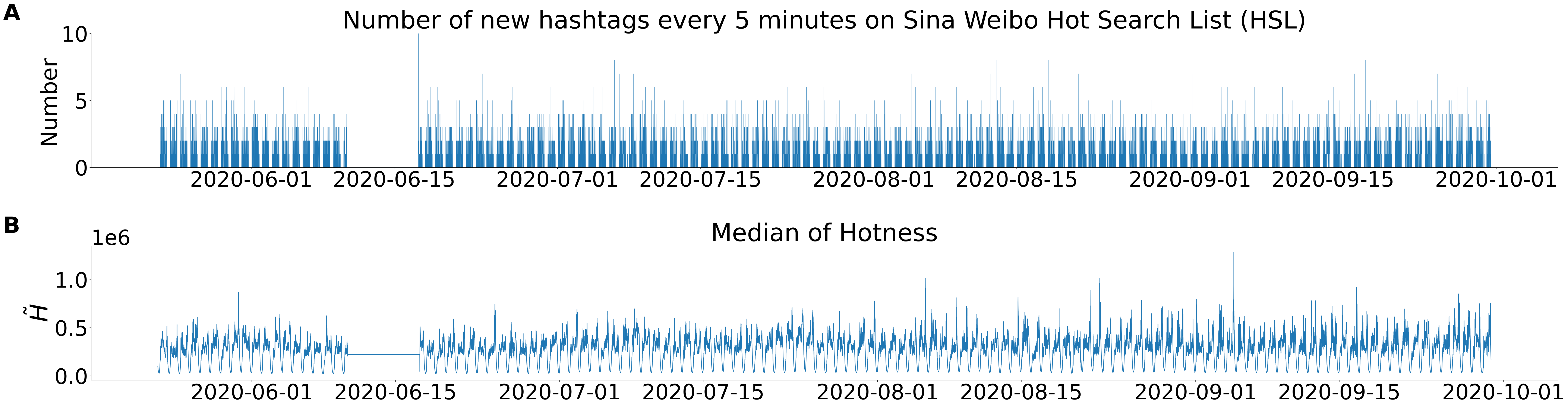}
    \end{center}
    \caption{\textbf{Circadian patterns of the Sina Weibo Hot Search List (HSL).} \textbf{(A)} Increment of number of new hashtags per $\Delta t=$ 5 minutes on the HSL during the observation period from 22 May 2020 to 29 September 2020. \textbf{(B)} Time series of the median of search volume index of all hashtags on the HSL at a timestamp, advertisement rank positions excluded. $\Tilde{H}$ represents the median value hotness $H$ of hashtags on Sina Weibo HSL at a timestamp. In both \textbf{(A)} and \textbf{(B)} the one-week gap due to the suspension of HSL by the cyberspace authority of China is visible. \textcolor{black}{An enlarged part of \bf{(A)} is in \nameref{S1_Appendix}.}} 
    
    \label{fig:fig1}
\end{figure}

Human actions are largely influenced by the circadian rhythm and so are online activities. Figure \ref{fig:fig1}A shows the increment of the number of hashtags per $\Delta t=$ 5 minutes interval clearly demonstrating the cyclic structure during the observation period from 22 May 2020 to 29 September 2020, except for a short interruption in June 2020. Similarly in Fig. \ref{fig:fig1}B, the median search volume index of hashtags on the HSL at a timestamp rises and decays in a periodic fashion. The missing of data for one-week in June 2020 is observed in both Fig. \ref{fig:fig1}A and Fig. \ref{fig:fig1}B, which results from the suspension of HSL by the cyberspace authority of China due to Weibo's interference with online communication \cite{suspension}. 

\subsection*{\textbf{Rank trajectory clustering}}

A successful hashtag $i$ stays on the HSL between the time instants $t_i$, when it appears on the list, until $T_i$, when it finally disappears from it defining the duration $d_i=T_i-t_i$. Some hashtags stay on the list for very short time ($d_i < 10$ minutes), while some others stay for many hours. The rank of a hashtag $i$ follows a trajectory $r_i(t)$. Some hashtags' trajectories first \textcolor{black}{go to higher ranks (smaller rank numbers) and then drop}, some go \textcolor{black}{higher and lower and higher} again, there are also cases that hashtag's trajectory goes \textcolor{black}{higher} and then it disappears. Also, the speed change of the trajectories is variable, resulting in a multitude of shapes of rank trajectories. 

\begin{figure}[htbp]
    \begin{center}
      \includegraphics[scale=0.23]{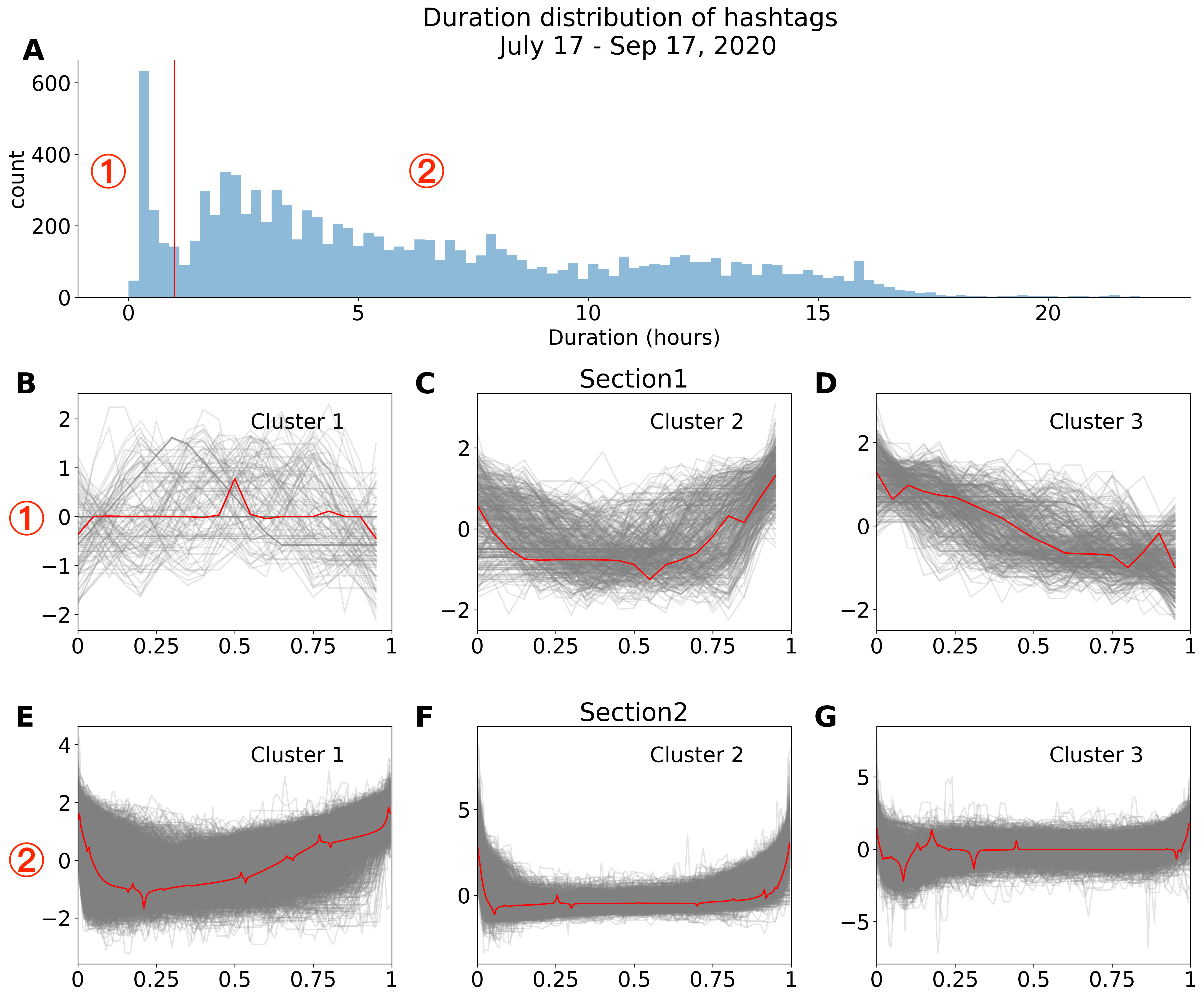} 
    \end{center}
    
    \caption {\textbf{Clustering patterns of hashtag rank trajectories on the Sina Weibo HSL.} \textbf{(A)} Distribution of hashtag duration on the HSL, divided into two sections based on local minima at 1 hour. Results of k-means clustering with 3 clusters in each section for time series data are shown, metric is \textcolor{black}{Dynamic Time Warping (DTW)} distance, y-axis is normalized to the mean and the standard deviation and the x-axis by $d_i$. \textbf{(B)}, \textbf{(C)}, \textbf{(D)} correspond to duration interval from 0 to 1 hour (Section 1). \textbf{(E)}, \textbf{(F)}, \textbf{(G)} correspond to duration interval larger than 1 hour (Section 2). Red curves depict clustering centers \textcolor{black}{(centroid) \cite{barycenter}, computed as the barycenters \cite{dtw} with respect to DTW. (We performed the clustering also with 4 clusters for both categories, see \nameref{S1_Appendix}.)}.} 
    
    \label{fig:fig2}
\end{figure}

The duration distribution of hashtags in the observation period is shown in Fig. \ref{fig:fig2}A. \textcolor{black}{We observe a sharp peak for hashtags with short duration and two less pronounced peaks.}
The vertical red line at the local minimum of 1 hour separates the duration distribution into two sections, section 1 and 2, respectively. The individual rows in Fig. \ref{fig:fig2} correspond to the clustering of the rank trajectories on each of the separated sections: Section 1 (B,C,D) and Section 2 (E,F,G).  
Even for hashtags with short duration on the HSL (Section 1) it is worth categorizing the rank trajectories. In most cases the rank does not change much during the lifetime $d_i$ (see Figs.~\ref{fig:fig2}B and C) and remains \textcolor{black}{at low ranks (large rank numbers)}, however, as shown in Fig. \ref{fig:fig2}D, some ranks of the hashtags exhibit a clear directional motion: they go to \textcolor{black}{higher ranks} and disappear from there. 
For the more expected rank trajectories shown 
from Fig. \ref{fig:fig2}E to Fig. \ref{fig:fig2}G, we also see some recognizable differences. 
Rank trajectories in Fig. \ref{fig:fig2}E first go \textcolor{black}{to higher ranks} and quickly go \textcolor{black}{to lower ranks} after hitting the top, without staying at a certain rank for a long time. 
Rank trajectories in Fig. \ref{fig:fig2}F first go \textcolor{black}{higher}, stay stable around the highest ranks with little fluctuation for a long time and then go down.
Rank trajectories in Fig. \ref{fig:fig2}G first go \textcolor{black}{higher}, with more fluctuations but never surpass the previous peak, then stay stable for a long time and finally go down \textcolor{black}{the ranks}. In the next Section we will show how the rank trajectory shapes are related to the time of the day the hashtags first appear on the HSL.

\subsection*{\textbf{Duration}}

\begin{figure}[htbp]
    \begin{center}
      \includegraphics[scale=0.33]{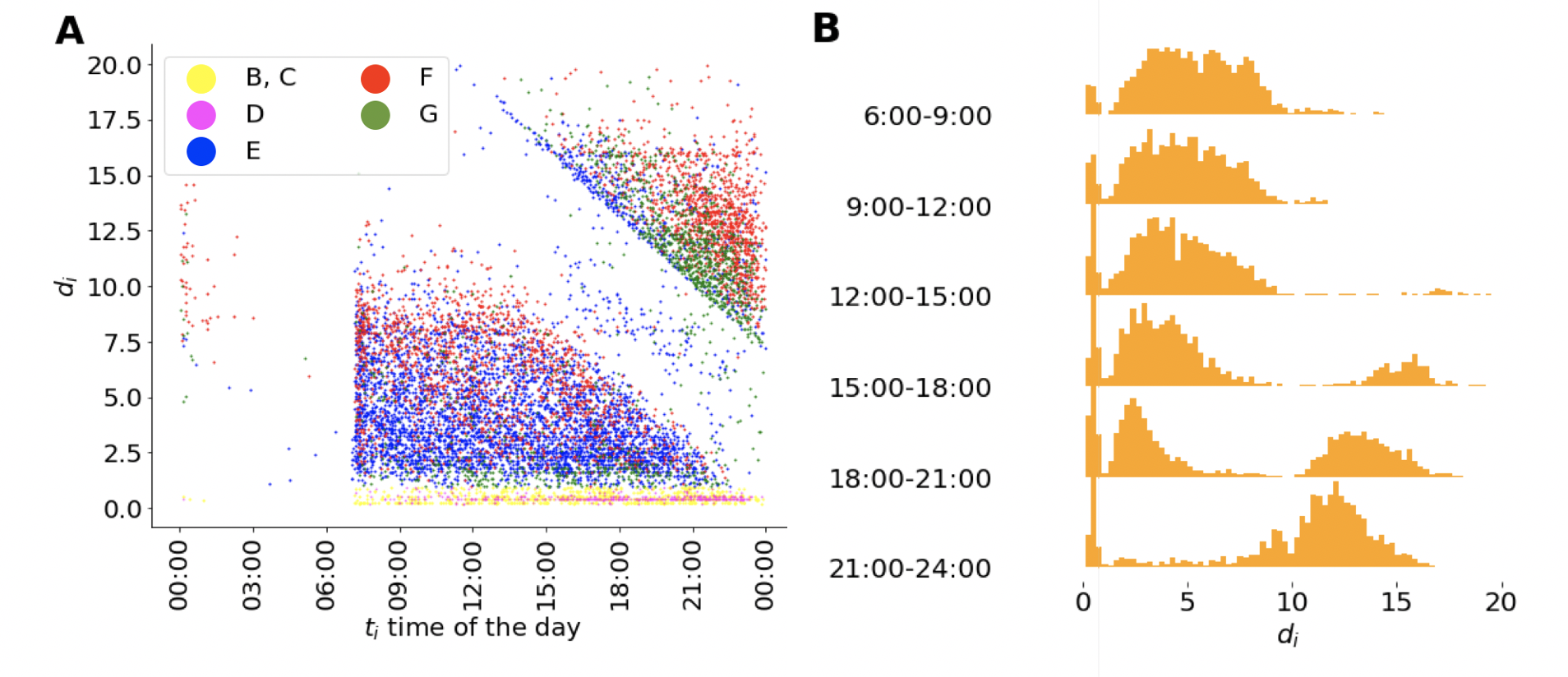} 
    \end{center}
    \caption{\textbf{Relationship between hashtags' duration on the HSL and the time $t_i$.} \textbf{(A)} Scatter plot of hashtags' duration on the HSL and the time of the day they first appear on the HSL. Each point is a hashtag, colored by the category it is clustered in  Fig. \ref{fig:fig2}. \textbf{(B)} Distribution of hashtags' duration on the HSL according to different time intervals during the day of first appearance on HSL.} 
    \label{fig:fig3}
\end{figure}

Figure \ref{fig:fig3}A is the $d_i$ vs $t_i(\bmod{\;24{\rm h})}$ scatter plot, i.e., it shows the durations of the hashtags vs the times of the day when they first appeared on the HSL, with each point representing a hashtag. Hashtags tend to appear on the HSL starting from around 7 a.m. till midnight. We can see clear shapes of lower-left and upper-right triangles, separated by a stripe in the middle with a low number of points inside. The lower boundary of the upper-right triangle is very sharp, while the upper boundary of the lower-left triangle is less so. There are data points within the stripe, but the density is much less compared with the data points inside the triangles and also if we compare it to the users' overall activity pattern (see \nameref{S1_Appendix}). The vertical distance between the triangle boundary lines is approximately 7 hours. The existence of these triangles suggests that the hashtags, which enter the HSL after 15 p.m. tend to either disappear from the HSL on the same day or stay on the HSL during the night and disappear after 7 a.m. the next day. This is presumably related to Sina Weibo working mode, already pointed out in previous studies~\cite{cui2022born}, namely that Sina Weibo practically stops \textcolor{black}{putting new hashtags onto HSL} between midnight and 7 a.m. If the ranking was automated following the formula Eq. \ref{eq:1}, the changes from day to night should not be that sharp and the circadian pattern should follow more or less that of the people's activity.

Figure \ref{fig:fig3}B shows the duration distribution of the hashtags as a decomposition of Fig. \ref{fig:fig2}A by binned starting values of the times of the day. For each time interval, the observed distribution is trimodal. As the start time of the day $t_i (\bmod{\;24{\rm h})}$ goes on, the density of hashtags in the third mode is increasing. 
In Fig. \ref{fig:fig3}A we see a low-density area at around 1 hour duration between the blue and the yellow dots, which corresponds to the minimum between sections 1 and 2 in Fig. \ref{fig:fig2}A. Accordingly, in the duration distribution plot shown in Fig. \ref{fig:fig3}B, a peak is observed for hashtags with duration shorter than 1 hour. Within this stripe \textcolor{black}{in Fig. \ref{fig:fig3}A,} there is an accumulation of pink dots corresponding to trajectories of category D, with a unique shape, namely starting at low rank and ending at a high one within a short period of time. In most other cases the more expected shape\textcolor{black}{s are} observed, namely starting and ending from low rank\textcolor{black}{s} and having in between some higher rank\textcolor{black}{s}. \textcolor{black}{Simple categorization of hashtags from each of the clusters based on semantic meaning does not point toward the relationship between the rank trajectories and the contents of the hashtag, as shown in \nameref{S1_Appendix} Table S1.  Later we will show that such an analysis for specific, anchoring ranks reveal systematic deviations from average behavior.} 

How are the shapes of the rank trajectories related to the time of the day the hashtags first appear on the HSL? Recall the Weibo working mode, if a hashtag's stay on the HSL is influenced by the night break, then it will automatically have a little-fluctuation period of at least seven hours, resulting in a rank trajectory shape similar to Fig. \ref{fig:fig2}F or the last part of Fig. \ref{fig:fig2}G, which we color in red and green respectively in Fig. \ref{fig:fig3}A. 
Hashtags in Fig. \ref{fig:fig2}F are born closer to midnight and further away from the hypotenuse of the upper-right triangle in Fig. \ref{fig:fig3}A. This is reasonable since hashtags entering HSL close to midnight are likely exposed to the stay on the HSL during the night break. Hashtags with shape in Fig. \ref{fig:fig2}G, however, are close to the hypotenuse boundary of the upper-right triangle in Fig. \ref{fig:fig3}A. One possible explanation is that although these hashtags' attention level is already in decreasing trend, their stay on the HSL are prolonged by the night break, so that when the next day begins, they are replaced by new hashtags and leave the list. 
The majority of hashtags with shape shown in Fig. \ref{fig:fig2}E are of shorter duration, located in the dense area of the lower-left triangle colored in blue in Fig. \ref{fig:fig3}A. 
The separation of the red and blue areas in Fig. \ref{fig:fig3}A lower-left triangle tells that hashtags which quickly go down after reaching their highest ranks on the HSL lack the ability to consistently attract public attention to maintain their positions on the list. In contrast, hashtags maintain relatively stable ranks (Fig. \ref{fig:fig2}F) stay longer times on the HSL, as Fig. \ref{fig:fig3}A lower-left red area suggests.     

\subsection*{\textbf{Ranking}}

\begin{figure}[htbp]
    \begin{center}
      \includegraphics[scale=0.42]{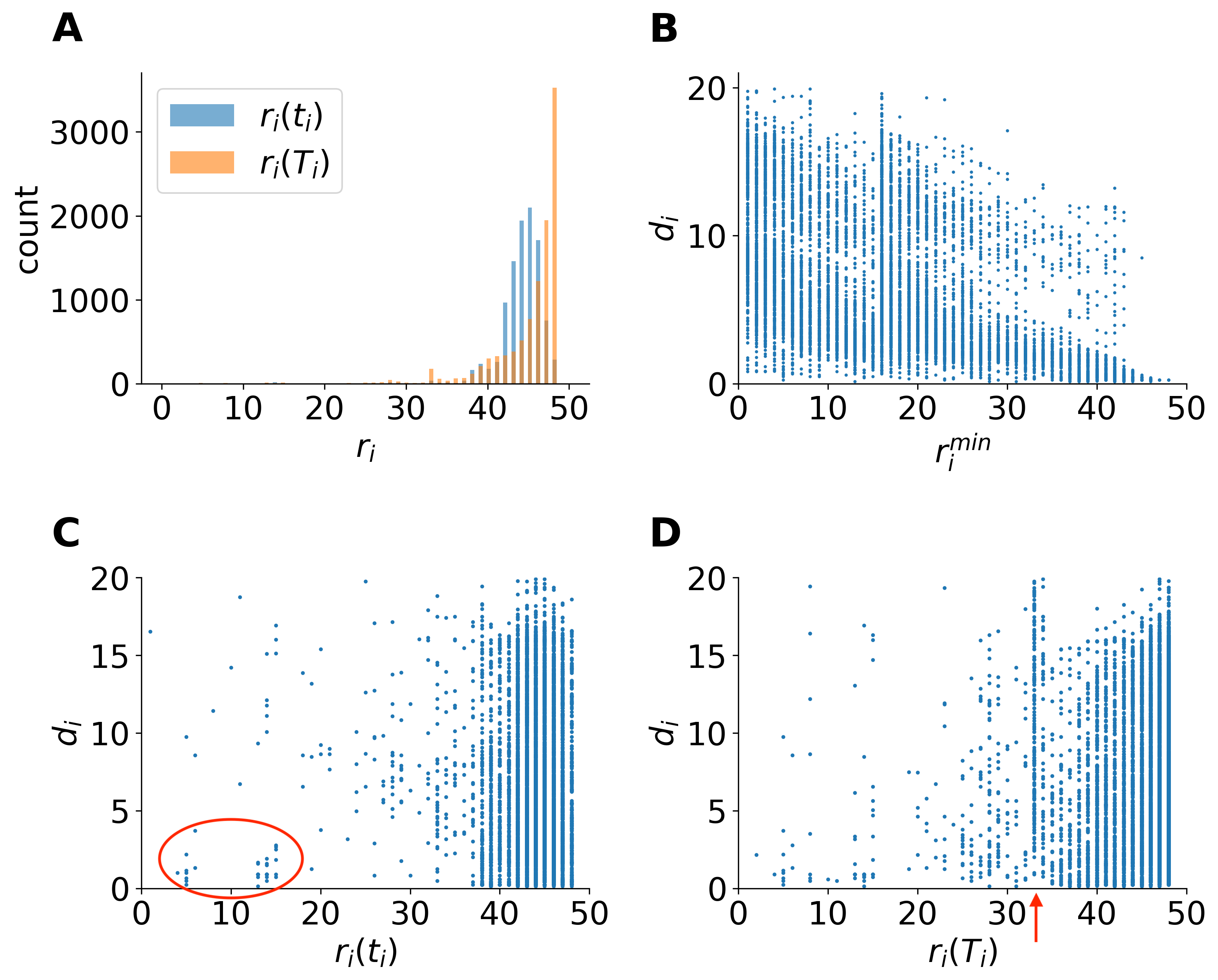}
    \end{center}
    \caption{\textbf{Ranking dynamics characterization of hashtags on the Sina Weibo HSL from 17 July 2020 to 17 Sep 2020.} \textbf{(A)} Distribution of $r_i(t_i)$ and $r_i(T_i)$.
    \textbf{(B)} Scatter plot of $r_i^{min}$ and $d_i$.
    \textbf{(C)} Scatter plot of $r_i(t_i)$ and $d_i$, \textcolor{black}{hashtags with high enter-rank and short duration are circled red. \textbf{(D)} Scatter plot of $r_i(T_i)$ and $d_i$, rank 33 marked by red arrow.} } 
    
    \label{fig:fig4} 
\end{figure}

\textcolor{black}{The popularity of a hashtag is reflected in its rank position and the duration it stays on the HSL.}
Figure \ref{fig:fig4}A shows the distribution of the enter-ranks $r_i(t_i)$ and leave-ranks $r_i(T_i)$ on the HSL. The majority of hashtags do not land on the HSL from the very bottom of the ranking list, instead they tend to enter at ranks 44 - 46 while they tend to leave from the bottom ranks. Figure \ref{fig:fig4}B shows the scatter plot of the highest rank of the hashtags and their duration on the HSL. The duration exhibits a bimodal pattern with a sudden jump at rank 16, and then it decreases. Figure \ref{fig:fig4}C shows the relationship between the hashtags' enter-ranks on the HSL and their corresponding duration on the HSL. 
\textcolor{black}{As known from ranking dynamics~\cite{iniguez2022dynamics}, items at high ranks stay longer, thus hashtags at higher ranks are more stable and stay for longer hours on the HSL,
so that it is strange for hashtags entering HSL at a high rank and only stay for a short duration, as marked in the red circle in Fig. \ref{fig:fig4}C. We found those hashtags are mostly related to celebrities, games, or TV programs, see \nameref{S1_Appendix}.  
In terms of the leave-rank and duration as shown in Fig. \ref{fig:fig4}D, rank 33 show a clear statistical deviation from the rest of the ranks, as marked by a red arrow. Surprisingly, we found the majority of those hashtags that leave HSL at rank 33 are related to international politics, to name a few, \#UK suspends extradition treaty with Hong Kong\# (\#\begin{CJK*}{UTF8}{gbsn}英国暂停与香港间的引渡条约\end{CJK*}\#),
\#The United States announced sanctions against 24 Chinese companies involved in the construction of islands in the South China Sea\#
(\#\begin{CJK*}{UTF8}{gbsn}美宣布制裁24家参与南海建岛中企\end{CJK*}\#), 
\#Russian foreign minister says won't reject 5G cooperation with China\#
(\#\begin{CJK*}{UTF8}{gbsn}俄外长称不会拒绝与中国开展5G合作\end{CJK*}\#). See \nameref{S1_Appendix} for the whole list of hashtags with leave-rank 33 together with their translation and examples of rank trajectories. }

\begin{figure}[htbp]
    \begin{center}
      \includegraphics[scale=0.27]{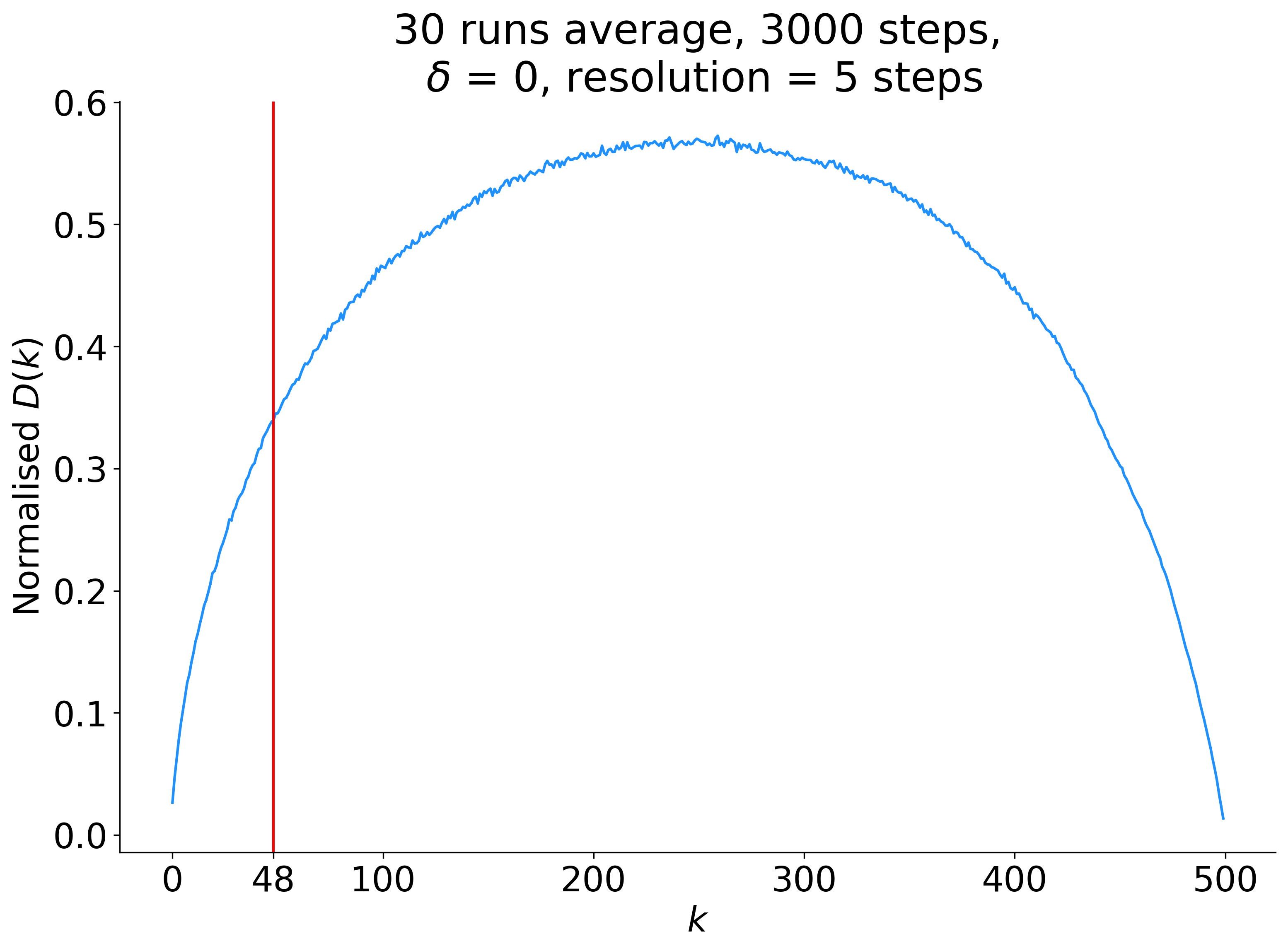}
    \end{center}
    \caption{\textbf{Parabola shaped normalized rank diversity in a model closed system of size 500.} The top L=48 can be considered as an open system.} 
    \label{fig:newfig} 
\end{figure}

As mentioned in the Introduction, spontaneously evolving ranking dynamics have typical rank diversity patterns~\cite{morales2016generic, iniguez2022dynamics, morales2018rank}. 
\textcolor{black}{A simple diffusional ranking model shows them clearly. Let us take a system of $N$ elements (for the hashtags), each element has a random initial score of values within $(0,1)$, and rank these elements from top to bottom based on the scores.  Let $r_i$ and $s_i$ denote the rank and the score of the $i$-th element, respectively. The scores change in time and that causes the rank movement of the elements. An element is randomly selected, 1 is added to its score and the ranking is changed if necessary. 
After sufficiently averaging the rank diversity, the curve shape is smoothed. Depending on whether the system is closed or open, the rank diversity curve shows a parabola shape or a portion of it, as shown in Fig. \ref{fig:newfig}.}


\subsection*{\textbf{Ranking dynamics in relation to prehistory}} 

\begin{figure}[htbp]
    \begin{center}
      \includegraphics[scale=0.4]{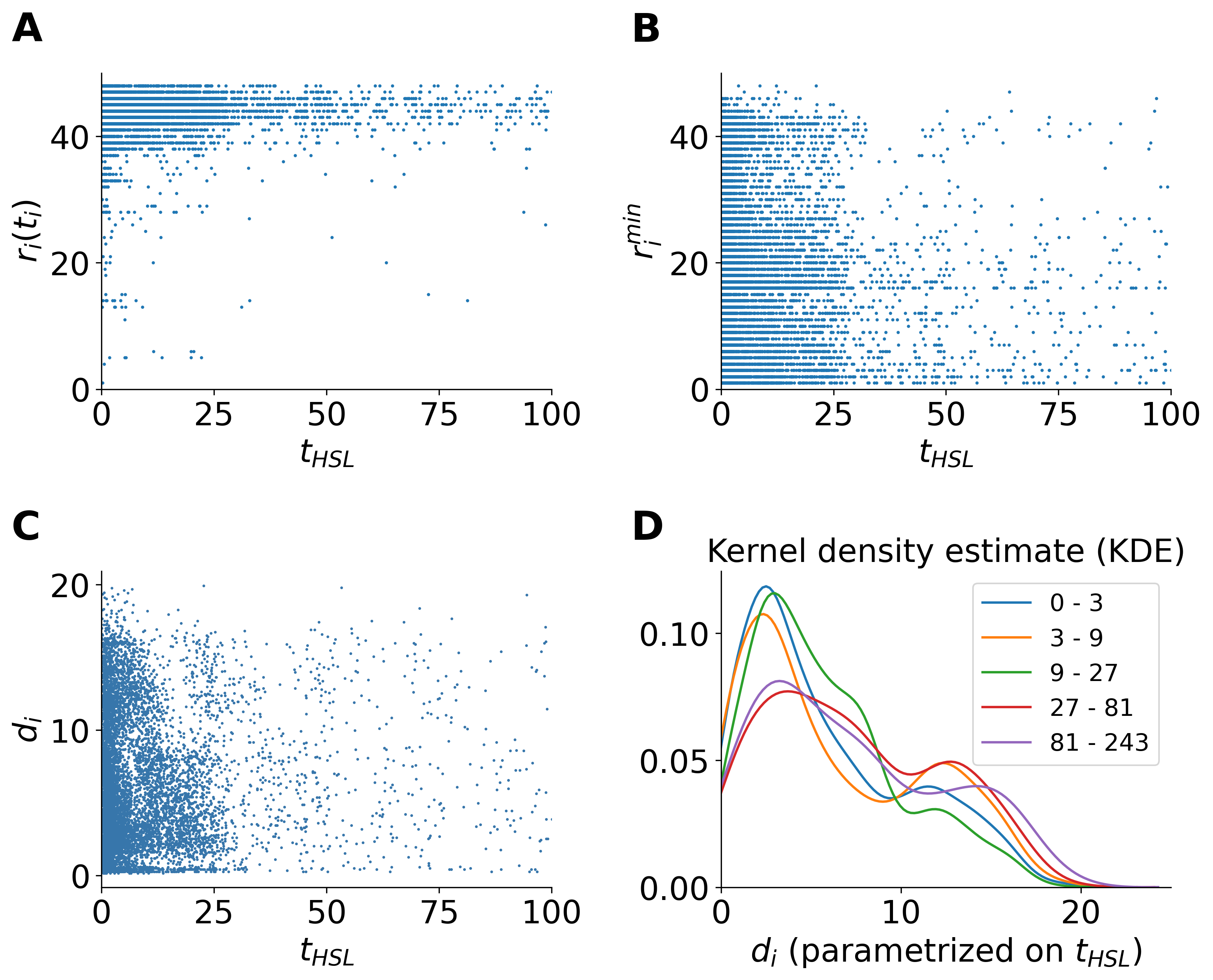}
    \end{center}
    \caption{\textbf{Prehistory length $t_{HSL}$, enter-ranks $r_i(t_i)$, \textcolor{black}{the highest rank $r_i^{min}$,} and duration $d_i$ of hashtags on the Sina Weibo HSL.} \textbf{(A)} The relationship between the hashtags' prehistory time length and the ranks they first enter on the HSL. 
    \textcolor{black}{\textbf{(B)} The relationship between the hashtags' prehistory time length and the highest rank during stay on the HSL.}
    \textbf{(C)} The relationship between the hashtags' prehistory time length and the duration they stay on the HSL. 
    \textbf{(D)} Parameterized probability density function of the hashtag duration on the HSL by prehistory time length, using kernel density estimation (KDE)~\cite{terrell1992variable}, with the parameter bw = ``scott"~\cite{scott2015multivariate}.} 
    \label{fig:fig5} 
\end{figure}

Before the hashtags gain enough popularity and land on the HSL, they go through different propagation routes during their prehistory. The time length of the prehistories $t_{HSL}$ differ for different hashtags~\cite{cui2022born}. Some hashtags get to the HSL in very short time after birth, while others take longer. Figure \ref{fig:fig5} shows the relationship between $t_{HSL}$ of the hashtags, the ranks they enter the HSL $r_i(t_i)$, \textcolor{black}{the highest rank $r_i^{min}$,} and the duration $d_i$ of their stay on the HSL. As shown in Fig. \ref{fig:fig5}A, in accordance with Fig. \ref{fig:fig4}A, the majority of hashtags enter the HSL at a low rank peaking around 45. Some hashtags enter the HSL at higher ranks, however, as the prehistory gets longer, the chance the hashtag enters the HSL from a high rank is less likely. \textcolor{black}{Figure \ref{fig:fig5}B suggests longer prehistory means in most cases higher top ranks ($r^{min}<20$). As shown in the previous work~\cite{cui2022born}, the hashtags about stars are over-represented in this category.}  As for the properties of hashtag duration on the HSL shown in Fig. \ref{fig:fig5}\textcolor{black}{C} and Fig. \ref{fig:fig5}D, the duration against prehistory length exhibits bimodal distribution. As the prehistory length increases, the first peak drops and the second peak rises. The bimodality similar to results shown in Fig. \ref{fig:fig3}, is influenced by the Weibo circadian working mode.

\subsection*{\textbf{Anchor effect}}
The dynamics of popularity as captured in the HSL should be sensitive to the actual trends and reflect the users' overall activity patterns. The individual rank trajectories show fluctuations but after averaging one would expect smooth behaviors. However, when studying the characteristics of the hashtags' rank dynamics on HSL, like the rank diversity we bumped into strange behaviors which we interpret as indications of interventions by the service provider.

\textcolor{black}{Here we generalize the ranking model introduced in Section Ranking to incorprate the anchoring effect} to simulate the dynamics of the hashtag ranking anomalies on the HSL. 
The idea of the anchor is the following: Set an anchor at position $A$. For hashtags whose $r < A$, it is difficult to go down the ranking list; for hashtags whose $r > A$, it is difficult to go \textcolor{black}{to higher ranks} (note that high rank means low $r$ value). The anchor represents a barrier characterized by an increment $\delta$. Let $\varphi(r_i)=i$ denote the selection of the element at a given rank at an instant of time.

The procedure of ranking at each step is shown below. Randomly pick one element $j$ and $s^{\rm new}_j=s_j+1$. There are three possibilities:\\
(a) $r_j < A$. Update the top $A-1$ ranks, no change of the anchor element.\\
(b) $r_j = A$. If $k=\varphi(A-1)$ and $s^{\rm new}_j > s_k + \delta$, update the top $A$ rank. Otherwise, no change of ranks.\\
(c) $r_j > A$. If $\ell =\varphi(A)$ and $s^{\rm new}_j>s_\ell+\delta$, old anchor rank drops to $A+1$, update the top $A+1$ ranks. Otherwise, update ranks lower than $A$, no change of the anchor element.

We simulate a system with 500 elements and take the top $L=48$ ranks to approximate an open system.  

\begin{figure}[htbp]
    \begin{center}
      \includegraphics[scale=0.45]{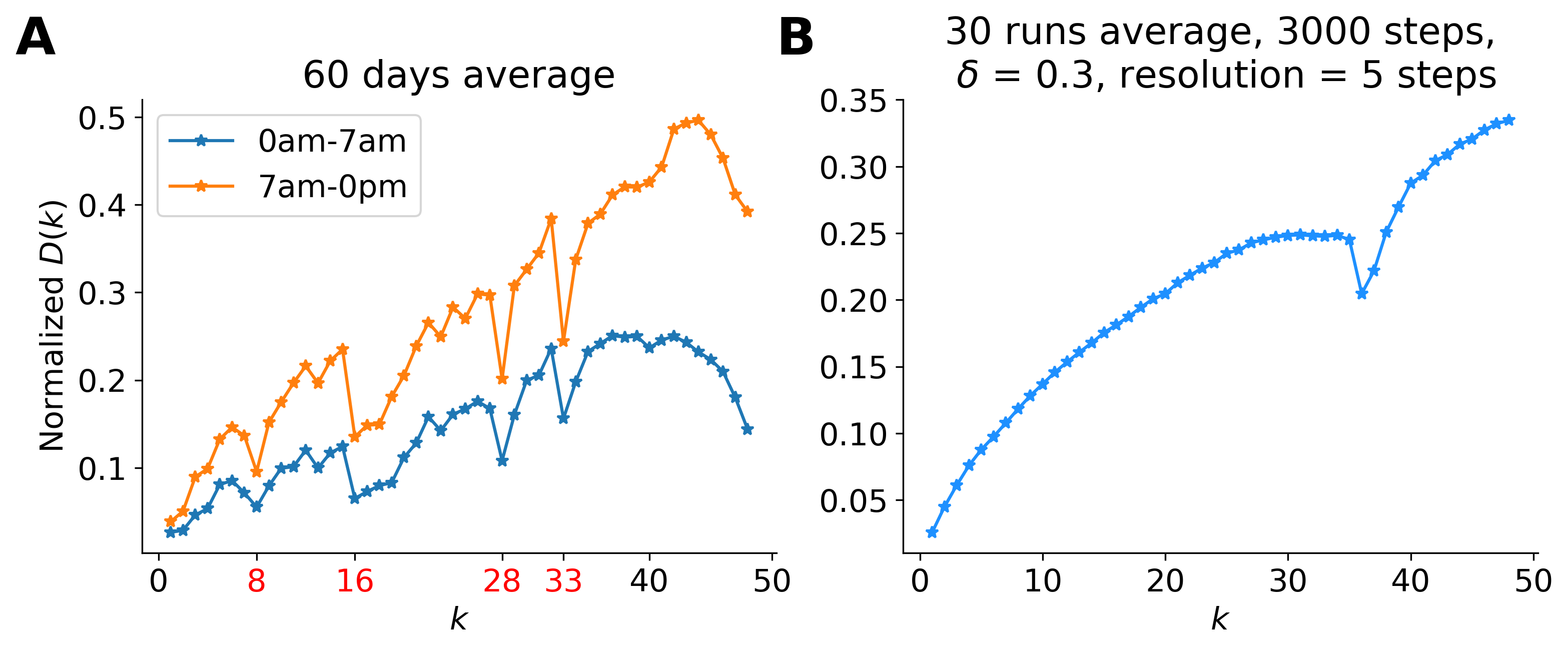}
    \end{center}
    \caption{\textbf{Rank dynamics comparison between empirical data and a ranking model with anchoring.} \textcolor{black}{\textbf{(A)} Empirical rank diversity separated for day (upper line) and night (lower line). The sudden drops are at ranks 8, 16, 28, and 33. \textbf{(B)} Simulated rank diversity with the anchor effect.}} 
    \label{fig:fig6}
\end{figure}

The rank diversity of a non-intervened system has parabola-shape\textcolor{black}{, see Fig. \ref{fig:newfig}}. The intervention produces a deep valley at the anchor\textcolor{black}{ing} position, very similar to those observed in the measured curves in Fig. \ref{fig:fig6} which shows the comparison between the real data and our model with anchoring. 
The difference between the behavior during the night and day is apparent: The former is more likely to the closed systems' characteristics with reduced activity while the latter is closer to the open systems' features although the trend around rank 44 turns down, probably due to the fact that the hashtags' enter-ranks $r_i(t_i)$ is shifted to the left as shown in Fig. \ref{fig:fig4}A. 
At certain positions (ranks 8, 16, 28, and 33) there are large drops in the values of the function, indicating intervention by ``anchoring" hashtags at these specific ranks. \textcolor{black}{With the simple model, reproducing qualitatively the effect, we support the assumption that the observed anomalies in the ranking functions are due to intervention. }


\begin{figure}[htbp]
    \begin{center}
      \includegraphics[scale=0.24]{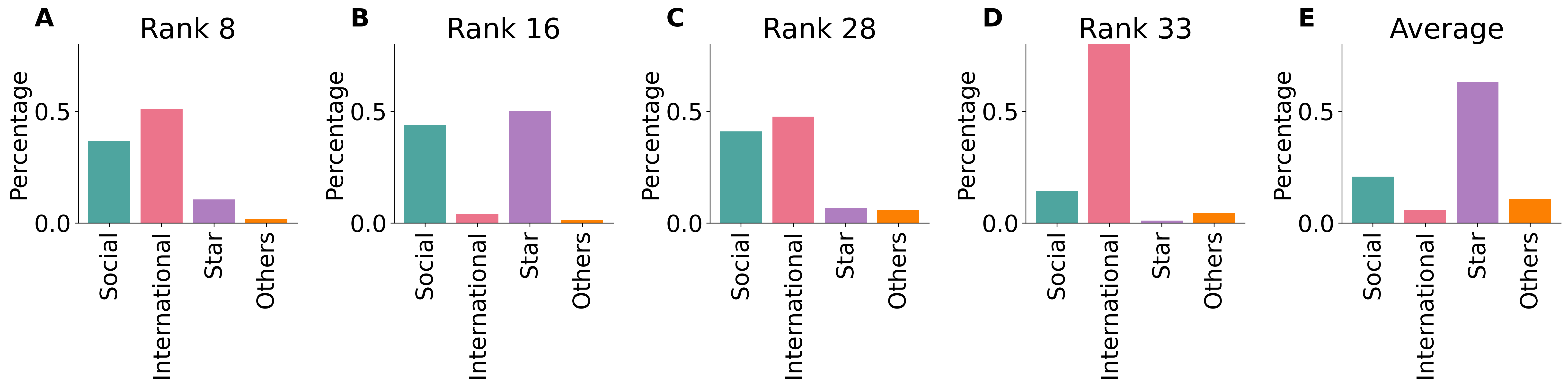}
    \end{center}
    \caption{\textcolor{black}{\textbf{Categorized proportion of hashtags that have stayed at certain ranks on HSL for longer than 2 hours.} \textbf{(A)(B)(C)(D)} show the content distribution of hashtags at ranks 8, 16, 28, and 33 respectively, corresponding to the sudden drops in Fig. \ref{fig:fig6}A. \textbf{(E)} Averaged proportion of hashtags by content category at ranks 5, 12, 21, 25, 30, 37.}}
    \label{fig:fig8}
\end{figure}

\textcolor{black}{To understand the background why some hashtags get anchored, we classified the hashtags that have stayed at the anchoring ranks for longer than 2 hours into four categories based on semantic meaning (see \nameref{S1_Appendix} SI3). Figure \ref{fig:fig8}(A)(B)(C)(D) show the proportion of such hashtags by category at each of the anchoring ranks 8, 16, 28, and 33, respectively. Comparing with Fig. \ref{fig:fig8}E, where the percentages are the average of each categories at six non-anchoring ranks (5, 12, 21, 25, 30, 37), ranks 8 (Fig. \ref{fig:fig8}A), 28 (Fig. \ref{fig:fig8}C), and 33 (Fig. \ref{fig:fig8}D) clearly have a large promoted proportion of International hashtags where the majority are related to international politics. Social hashtags also have a larger proportion at anchoring ranks except for rank 33.  
}

\section*{Discussion}

Public attention is precious and it is nowadays largely dependent on online social media, therefore it is of great interest to understand the dynamics governing popularity on such platforms. Considerable effort has been devoted to this task on Twitter ~\cite{wu2007novelty,eom2015twitter,annamoradnejad2019comprehensive,asur2011trends} and some results are also known on Sina Weibo~\cite{fan2015topic,yu2015trend,cui2022born}. 
In order to attract attention, people, companies, and political actors are tempted to make use of hidden manipulations besides well known tools of direct advertisements or propaganda~\cite{zhang2016twitter,stafford2013evaluation,bamman2012censorship,cui2021attention,cui2022born}. 
Thus popularity can emerge spontaneously via collective attention from online users who are genuinely interested in a topic and form trends, quantified and captured by the algorithm of the platform, or trends emerge from intervention by the platform provider motivated by financial or other interests. (It should be noted that ``collective attention" may also be influenced, e.g., by spamming~\cite{stafford2013evaluation} or coordinated campaigns~\cite{yu2015trend,pina2022coordinated}.)

In this paper, we studied the attention dynamics of trending hashtags on the Sina Weibo Hot Search List by using various measures of ranking dynamics, like entering and leaving ranks as well as duration of hashtags on HSL, rank diversity, and categories of rank trajectories. The aim of the identification of regularities in the ranking dynamics was twofold: First, contribution to the quantitative characterization of the dynamics of public attention in order to better understand its mechanism, and second, finding signatures of \textcolor{black}{possible} interventions by the service provider. 

The duration of the hashtags on the HSL in relation to the time of the day they enter the list shows trimodality (Fig. \ref{fig:fig3}). This is related to the fact that the appearance of hashtags on the HSL have circadian patterns (Fig. \ref{fig:fig1}A). On the one hand, the pattern is caused  by the circadian rhythm of the users whose activities depend on the time of the day (see \nameref{S1_Appendix}), on the other hand it is imposed by the apparent working mode of Sina Weibo, which reduces the night-time flow of new hashtags to the HSL almost to zero level. The night break is reflected in the very low number of points in the stripe separating the two triangles in Fig. \ref{fig:fig3}A and in the particularly sharp upper boundary of this stripe. This seven hour gap has been shown to influence the prehistory of the successful hashtags~\cite{cui2022born} by contributing to the difference between shorter and longer prehistories and it creates a link between the behavior of the hashtags on the HSL and their prehistory (Fig. \ref{fig:fig5}). 



\textcolor{black}{The distinction between user daily posts volume shown in \nameref{S1_Appendix} Fig. S1 and the sharp day-night boundaries of daily patterns in \nameref{S1_Appendix} Fig. S2} is already an example that we are able to identify interventions by the service provider, that the ranking is not automated following a plain formula like Eq.~\ref{eq:1} but depends on human control. 
More importantly, we show an anchoring effect at some rank positions on the HSL, where rank diversity is suppressed as compared to the expected smooth behavior of this quantity. Using a simple ranking model we show how anchoring at some rank positions changes rank diversity. A further observation indicating intervention is that some hashtags on the HSL appear at high ranks and disappear in short time (Fig. \ref{fig:fig4}C), \textcolor{black}{we found these hashtags are mostly from the Star category, (see \nameref{S1_Appendix} for the categorization and the list of the hashtags).}  
Similarly, there are many hashtags that just stay on the HSL for short time which is shown in the first peak in Fig. \ref{fig:fig3}B. The fact that the peak is separated from the rest of the distribution is also likely be related to intervention.

\textcolor{black}{Our method cannot tell the origin of the interventions, whether they result directly from Weibo, pollution by bots, or internet water armies \cite{waterarmy} that could blur the picture of the natural activities originating from normal users. The fact that the irregularities occur at specific ranks, which would be difficult to target by external influence makes the intervention by the service provider more likely. The reason or motivation for the possible intervention is unknown, we can only make reasonable guesses based on statistical analysis, whether it is due to the government’s “promote positive contents” campaign, or the influence of social capital to promote certain advertisements, etc. 
For example, the proportion of International hashtags that have stayed for longer than 2 hours at the anchoring ranks (8, 28, 33) are much larger than the averaged value of several other non-anchoring ranks. In addition, we found that the hashtags leaving the HSL at rank 33 are mostly related to international news, predominantly of political nature. Moreover, politics related hashtags tend to have less fluctuations than non-politics hashtags (see \nameref{S1_Appendix} Fig. S10), implying possible political motivation. Ranks 8, 16, and 28 also see a higher proportion of anchored Social hashtags, this might be an implication of Weibo's social responsibility as an important news source of social issues for the public.}


Sina Weibo is the microblogging site with world-wide the largest number of active users, who are overwhelmingly Chinese speakers. While we believe that alone the size of Sina Weibo justifies focused study, we know that most of our results are idiosyncratic. However, this is true only in a narrow sense as our results provide general lessons. We demonstrated that studying the ranking dynamics in popularity lists is worth for several reasons. First, we uncovered relationships between ranking dynamics and the circadian pattern of user activity, also establishing a link to the prehistory of items getting to the ranking list. Moreover, we identified different trajectory categories on the list, which characterize different dynamic patterns of popularity.  Finally, and most importantly, we showed, how pinpointing anomalies in ranking statistics can be used to identify interventions by the service provider. As service providers have financial interests and may be under political pressure, objectivity of the ranking lists and its truth content can be questioned. 

\textcolor{black}{As the platform algorithms may change from time to time, it is challenging to keep track 
of the interventions, 
as they can be detrimental being a possible tool of online mass manipulation. Thus, similar}
to the fight against fake news, the fight against manipulation of public attention is in the interest of the society and it also needs the tools of detecting interventions. \textcolor{black}{Our studies give important reference not only in terms of intervention detection on social media, but also for other research disciplines, such as communication science, journalism, political science, to investigate in further details the specific messages and different aspects of online political contents, and learn more about the motivations of such interventions.}


\section*{Supporting information}


\paragraph*{S1 Appendix.}
\label{S1_Appendix} {\bf Supplementary Material to the manuscript.}

\section*{Acknowledgments}

We are grateful to Gerardo I\~niguez for his valuable advice.

\nolinenumbers

%
%
%



\section*{Supplementary material}

**Supporting Information to:**

**Competition for popularity and interventions on a Chinese microblogging site**

Hao Cui[1] and János Kertész[1]*

[1]Department of Network and Data Science, Central European University, Quellenstrasse 51, A-1100 Vienna, Austria

*Correspondence: kerteszj@ceu.edu

## SI1. Daily posts volume on Weibo

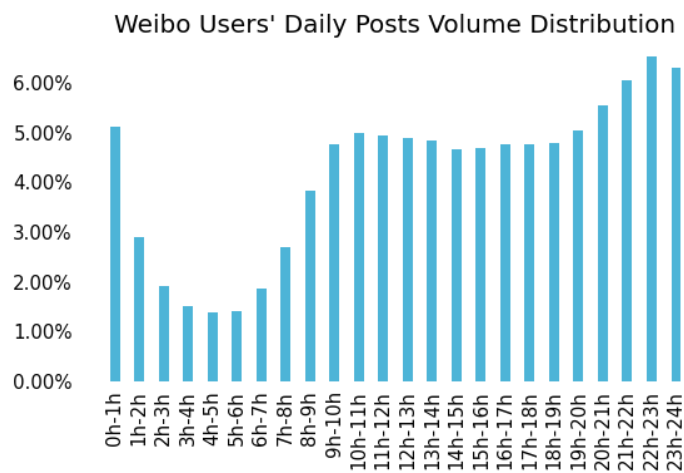

**Figure S1**. Distribution of Weibo users' daily posts volume according to Weibo User Development Report [1].

## SI2. Enlarged part of Figure 1A

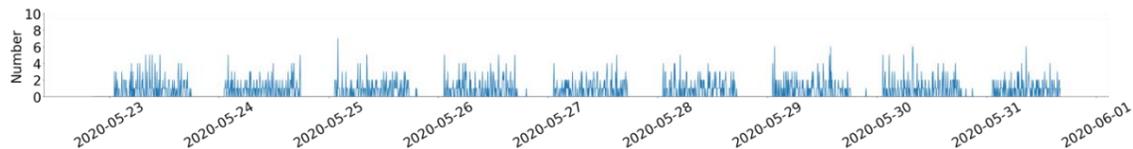

**Figure S2**. Number of new hashtags every 5 minutes on Sina Weibo Hot Search List (HSL). An enlarged part of Fig. 1A with clear cyclic structure.

## SI3. Categorization of hashtags in each cluster and examples of rank trajectories

**Table S1.** Number of hashtags by category in sampled clusters of size 100

|                    | Stars | Social | International | Others |
|--------------------|-------|--------|--------------|--------|
| Section 1 Cluster 1 | 34    | 46     | 12           | 8      |
| Section 1 Cluster 2 | 43    | 35     | 8            | 14     |

| Section 1 Cluster 3 | 47 | 36 | 11 | 6 |
| Section 2 Cluster 1 | 42 | 40 | 12 | 6 |
| Section 2 Cluster 2 | 57 | 28 | 8 | 7 |
| Section 2 Cluster 3 | 44 | 30 | 12 | 14 |

We took random samples of size 100 for each of the clusters and classified the hashtags into 4 categories based on their semantic meaning by human judgement: Stars, Social, International, and Others. The Stars category consists of movie/sports stars, singers, idols, celebrities as well as the TV programs/movies and events that they participate in. The Social category consists of hashtags that are related to social accidents, crimes, natural disasters, and other events that are related to social life. The International category consists of hashtags whose content is related to news of regions and countries outside of mainland China. The Others category consists of the rest of the hashtags that fall into none of the above categories. There are not huge differences between the proportions of each category in each cluster. The Star category and the Social category occupies around 80% of the total sample. The following figures show the rank trajectory examples of hashtags in each cluster. The random samples of each cluster are available in the GitHub repository [2].

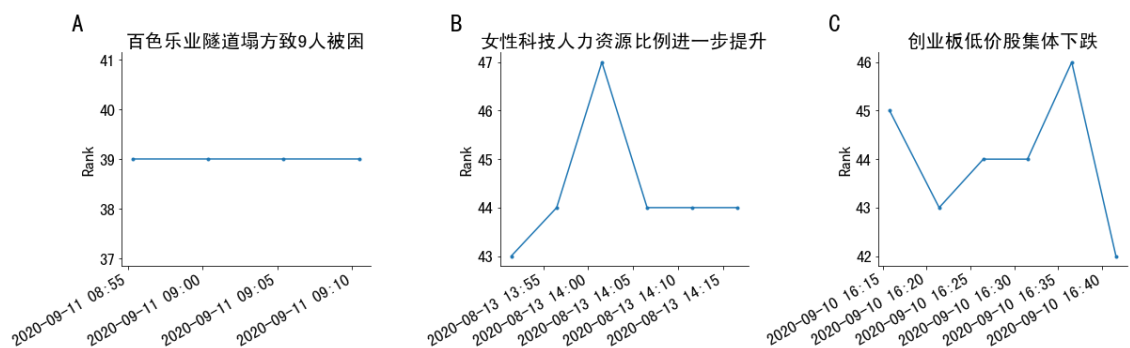

**Figure S3**. Examples of rank trajectories in Section 1 Cluster 1. (A) # Baise Leye tunnel collapse caused 9 people trapped# (#百色乐业隧道塌方致 9 人被困#) (B) (#The proportion of female science and technology human resources to further improve#) #女性科技人力资源比例进一步提升# (C) # GEM low-priced stocks fell collectively# (#创业板低价股集体下跌#)

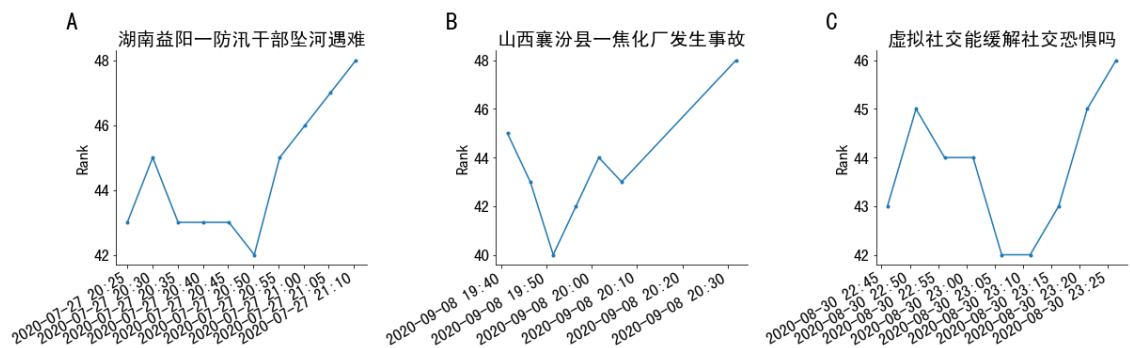

**Figure S4**. Examples of rank trajectories in Section 1 Cluster 2. (A) #Hunan Yiyang a flood control cadres fell into the river and died# (#湖南益阳一防汛干部坠河遇难#) (B) (#An accident occurred in a coking plant in Xiangfen County, Shanxi#) #山西襄汾县一焦化厂发生事故# (C) #Can virtual socializing ease social fears# (#虚拟社交能缓解社交恐惧吗#)

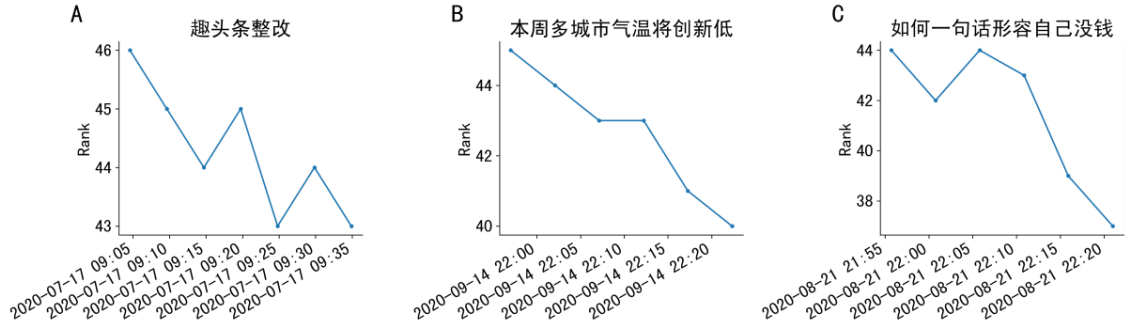

**Figure S5**. Examples of rank trajectories in Section 1 Cluster 3. (A) #Qutoutiao rectification# (#趣头条整改#) (B) (#Temperatures set to hit record lows in many cities this week#) #本周多城市气温将创新低# (C) #How to describe yourself having no money in one sentence# (#如何一句话形容自己没钱#)

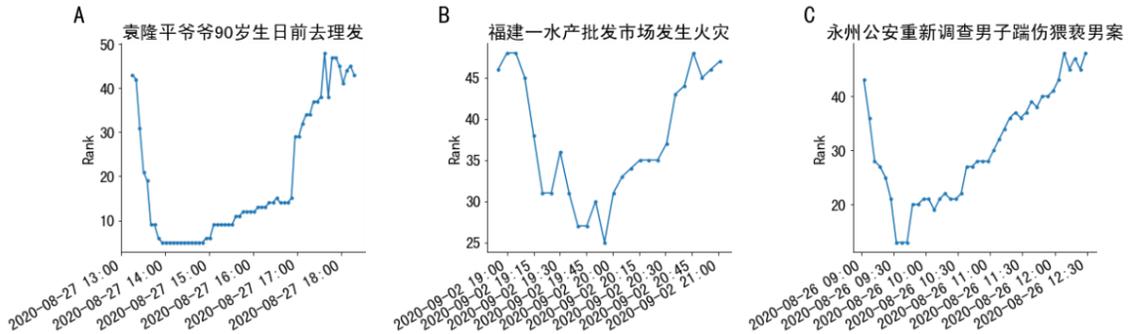

**Figure S6**. Examples of rank trajectories in Section 2 Cluster 1. (A) # Grandpa Yuan Longping went for a haircut before his 90th birthday# (#袁隆平爷爷 90 岁生日前去理发#) (B) (#Fire in a wholesale fish market in Fujian#) #福建一水产批发市场发生火灾# (C) # Yongzhou public security re-investigation of the man kicked and injured molesting man# (#永州公安重新调查男子踹伤猥亵男案#)

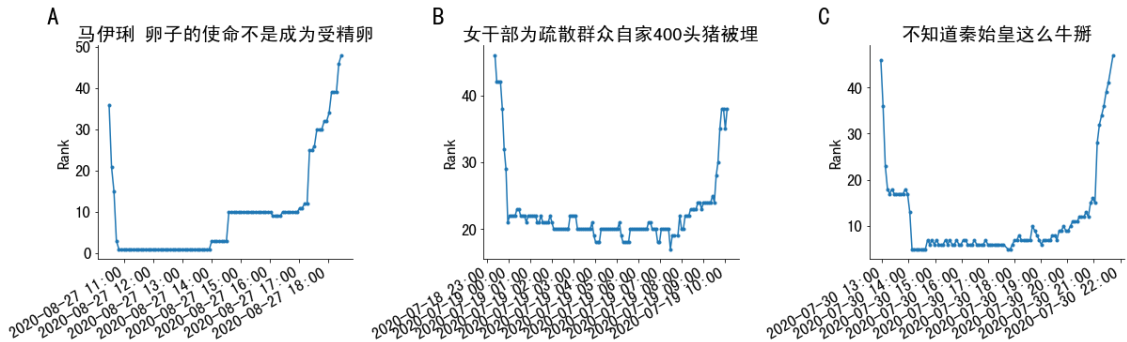

**Figure S7**. Examples of rank trajectories in Section 2 Cluster 2. (A) # Ma Yili The mission of the egg is not to become a fertilized egg# (#马伊琍 卵子的使命不是成为受精卵#) (B) (#Female cadres to evacuate the crowd, 400 pigs buried in own home#) #女干部为疏散群众自家 400 头猪被埋# (C) #Did not know Qin Shi Huang was so powerful# (#不知道秦始皇这么牛掰#)

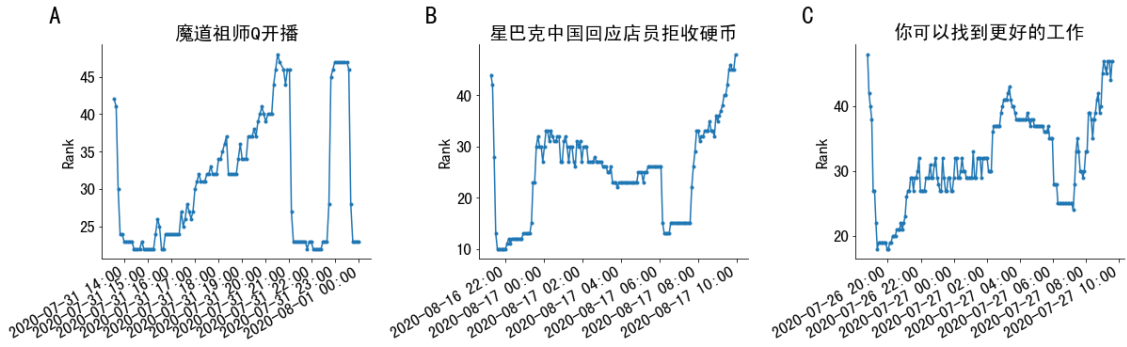

**Figure S8**. Examples of rank trajectories in Section 2 Cluster 3. (A) #Grandmaster of Demonic Cultivation Q start of broadcast# (#魔道祖师 Q 开播#) (B) (#Starbucks China responds to store clerk's refusal to accept coins#) #星巴克中国回应店员拒收硬币# (C) #You can find better jobs# (#你可以找到更好的工作#)

## SI4. Rank trajectory clustering patterns with different number of clusters

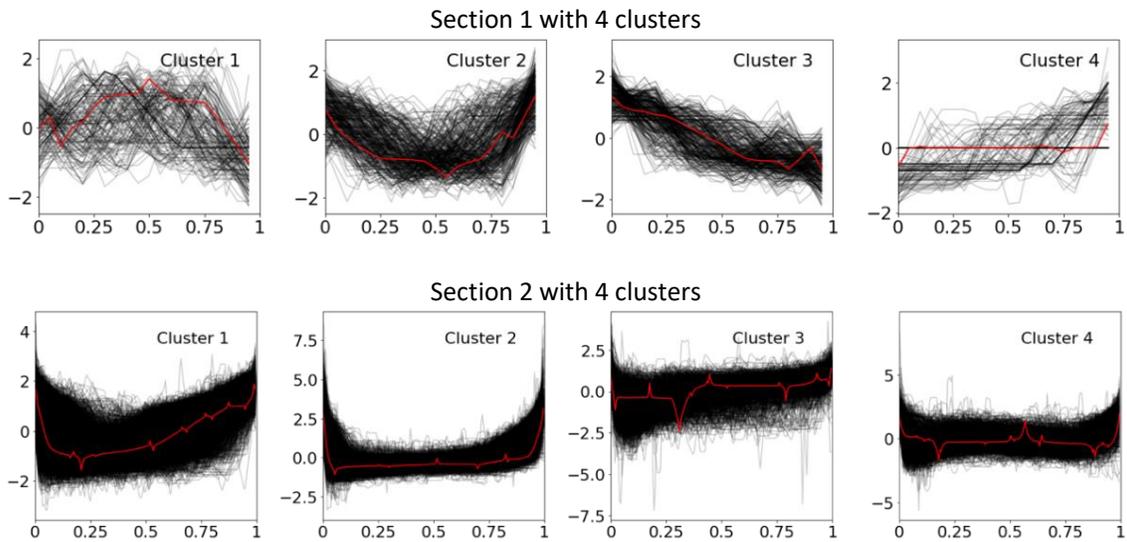

**Figure S9**. Rank trajectory clustering patterns of each section with different number of clusters.

## SI5. Hashtags with high enter-rank and short duration

**Table S2.** Hashtags with high enter-ranks and short duration on Weibo HSL

| Hashtags | Translation |
|---|---|
| 令人叫绝的明星神颜 | The amazing faces of the stars |
| 这届妻子不好惹 | Wives in this episode are not to be messed with |
| 李斯丹妮汉服版无价之姐 | Dany Lee Hanfu version of priceless sister |
| 百家电影官微集体营业 | Hundreds of official movie microblogs are open for business |
| 明星跨界有多拼 | How hard do stars cross the border |
| 街舞 3 队长 battle | Street dance 3 captain battle |
| 尹正太有梗了 | Yin Zheng is so gag |
| 潘晓婷台球教学 | Pan Xiaoting billiards teaching |
| 王者荣耀世冠神秘战队爆料 | Honor of Kings world championship mystery team break the news |
| 李荣浩套路终结者 | Li Ronghao trick terminator |
| 欧阳娜娜给薇娅采耳 | Ouyang Nana picks Viya's ear |
| 梁建章扮包拯带你游 | Liang Jianzhang dressed as Bao Zheng to take you on a tour |
| 云鹰飞将 QGFly 对线夫赖 | Cloud Eagle Flying General QGFly vs. Fly |
| 医生谢谢你 | Doctor thank you |
| 如何看出帅哥喜欢我 | How to see that handsome guy likes me |
| 女生玩游戏太上头了 | Girls play games too addicted |
| 陈一冰倒立涂口红 | Chen Yibing upside down applies lipstick |
| 健身素颜妆 | fitness vegan makeup |
| 秦昊宠女儿 | Qin Hao spoils his daughter |
| 和平精英 PEL 冠军 | Game for peace PEL champion |
| 足不出户云游王者 | Remain within doors roam king |
| 潮玩人类在哪里 | Where are the tide playing humans |
| 吴建豪灵魂拷问苏五口 | Wu Jianhao soul searching Suwukou |
| 雪梨苏芒谈年龄状态 | Cherie Su Mang talks about age status |
| 李汶翰壁咚张鹤伦 | Li Wenhan kabe-don Zhang Lunhe |
| 沈腾拒绝李汶翰 | Shen Teng rejects Li Wenhan |
| 被女玩家支配的恐惧 | Fear of being dominated by female gamers |
| 电竞为什么没有女职选手 | Why there are no female professional players in e-sports |
| 秦昊薇娅合唱 | Qin Hao Weiya chorus |
| 电子竞技莫得感情 | e-sports got no feelings |
| 雪梨杨天真合伙 | Cherie Yang Tianzhen partnership |

| 金晨郁可唯表情包 | Jin Chen Yu Kewei image macro |
| 镜头中的脱贫故事 | The story of poverty eradication in the lens |
| TS 对阵 DYG | TS vs DYG |

## SI6. Hashtags leave HSL at rank 33

The following table shows the list of the hashtags that have left the re-ranked Sina Weibo Hot Search List (HSL) at the rank 33, together with their English translations. Among 177 such hashtags, 138 of them are related to news of regions and countries outside of mainland China as shown in bold, and an overwhelming number of them are related to international politics. Among those domestic hashtags, a large majority are related to domestic regulations or politics. Example rank trajectories are shown in Fig. S10.

***Note:*** *the Weibo HSL we mention in this paper refers to the re-ranked HSL after removing the advertisements labeled with "Recommendation (荐)".*

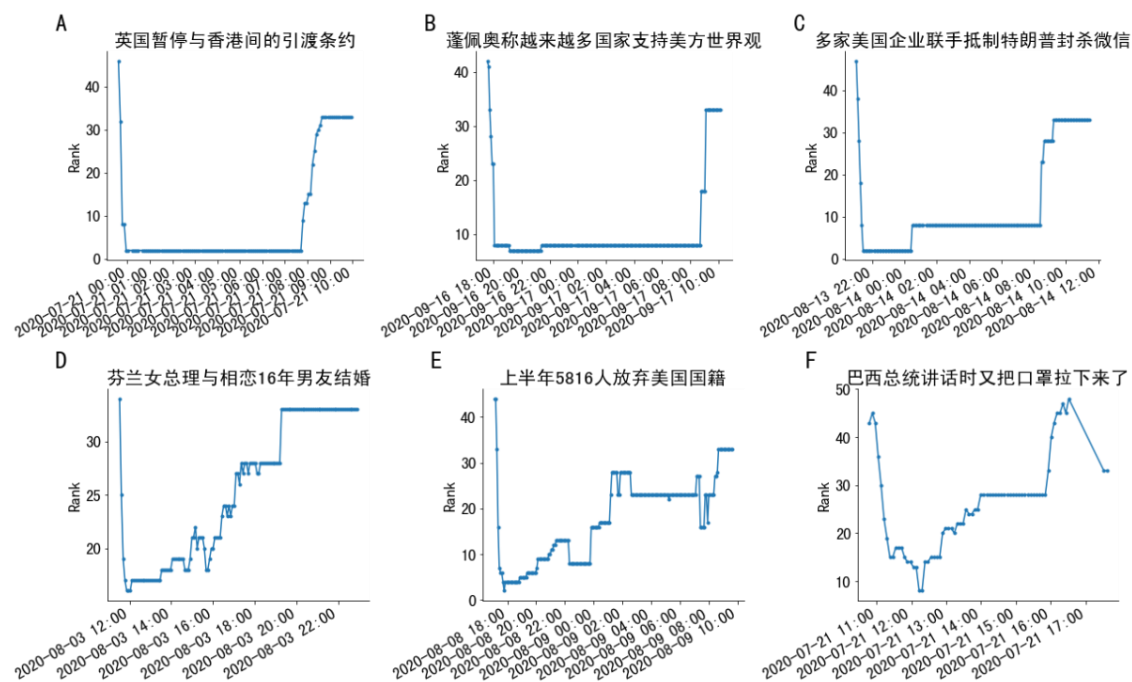

**Figure S10.** Examples of rank trajectories of hashtags leave HSL at rank 33 in the International category. Hashtags related to international politics have less fluctuations in their rank trajectories. (A) #UK suspends extradition treaty with Hong Kong# (#英国暂停与香港间的引渡条约#) (B) #Pompeo says more countries support US worldview# (#蓬佩奥称越来越多国家支持美方世界观#) (C) #Many U.S. companies join forces to resist Trump's blocking of WeChat# (#多家美国企业联手抵制特朗普封杀微信#) (D) #Finnish female PM marries boyfriend of 16 years in love# (#芬兰女总理与相恋 16 年男友结婚#) (E) #5,816 people renounce U.S. citizenship in first half year# (#上半年 5816 人放弃美国国籍#) (F) #Brazil's president pulled down the mask again while speaking# (#巴西总统讲话时又把口罩拉下来了#)

**Table S3.** List of hashtags that leave Weibo HSL at rank 33

| Hashtag | Translation |
| --- | --- |
| 31 省区市新增确诊 16 例 | 16 new cases confirmed in 31 provinces and cities |
| 长征五号火箭垂直转运至发射区 | Long March 5 rocket vertical transfer to the launch area |
| **扎克伯格连线福奇** | **Zuckerberg connected to Fauci** |
| 国务院联防联控机制联络组离鄂返京 | The State Council joint prevention and control mechanism liaison group left Hubei to Beijing |
| **特朗普称将恢复白宫每日疫情简报会** | **Trump says he will resume daily White House outbreak briefings** |
| **巴西总统讲话时又把口罩拉下来了** | **Brazil's president pulled down the mask again while speaking** |
| **外交部回应在永兴岛部署战斗机** | **Foreign Ministry responds to deployment of fighter jets in Yongxing Island** |
| **世卫反驳称蓬佩奥指责毫无根据** | **WHO counters that Pompeo's accusations are baseless** |
| **白俄总统确诊感染新冠病毒** | **Belarus president diagnosed with coronavirus** |
| **佩洛西要求美国会众议员佩戴口罩** | **Pelosi asks U.S. congressmen to wear masks** |
| 科技部明确论文数量不可与奖励挂钩 | Ministry of Science and Technology clarifies that the number of papers cannot be linked to awards |
| **外交部回应美国威胁巴西排斥华为** | **Foreign Ministry responds to US threat to reject Huawei in Brazil** |
| **国台办回应李登辉病亡** | **Taiwan Affairs Office responds to Lee Teng-hui's illness and death** |
| 钟南山成为共和国勋章建议人选 | Zhong Nanshan becomes recommended candidate for Order of the Republic |
| **外交部向 BBC 提出严正交涉** | **Foreign Ministry lodges stern representations with BBC** |
| **黎巴嫩环境部长辞职** | **Lebanon's environment minister resigns** |
| **港府全力支持中央政府对美反制** | **The Hong Kong government fully supports the central government's countermeasures against the US.** |
| **拜登提名美国副总统候选人** | **Biden nominates US vice presidential candidate** |
| **福奇质疑俄罗斯新冠疫苗** | **Fauci questions Russia's coronavirus vaccine** |
| 测谎结果不属于合法证据形式 | Polygraph results are not a legitimate form of evidence |
| 商务部发文称将开展数字人民币试点 | Ministry of Commerce issues document saying it will carry out digital yuan pilot |
| **外交部谴责美将孔子学院妖魔化污名化** | **Foreign Ministry condemns US demonization and stigmatization of Confucius Institute** |
| **香港警队回应英国禁止培训港警** | **Hong Kong police force responds to UK ban on training Hong Kong police** |
| **新西兰因疫情反弹决定推迟大选** | **New Zealand decides to postpone general election due to backlash against epidemic** |

| 外交部回应加拿大与 FBI 合谋搜查孟晚舟 | Foreign Ministry responds to Canada's conspiracy with FBI to search Meng Wanzhou |
|---|---|
| 商务部回应美封杀 TikTok | Commerce Ministry responds to US blocking TikTok |
| 班农申请无罪审判 | Bannon files for not guilty trial |
| 班农交 500 万美元保释金离开法院 | Bannon pays $5 million bail to leave court |
| 解放军和武警部队 120 多万人次抗洪 | PLA and armed police forces more than 1.2 million people to fight floods |
| 全球已有超 8 万个新冠病毒基因序列 | The world has more than 80,000 new coronavirus gene sequences |
| 特朗普政府或允许美企在中国使用微信 | Trump administration may allow U.S. companies to use WeChat in China |
| 王毅批美国制裁守法国家荒唐透顶 | Wang Yi criticizes US sanctions against law-abiding countries as absurd |
| 驻以色列使馆提醒中国公民防范新冠 | Embassy in Israel reminds Chinese citizens to guard against coronavirus |
| 中方候选人当选国际海洋法法庭法官 | Chinese candidate elected to International Tribunal for the Law of the Sea |
| 巴西总统长子确诊感染新冠肺炎 | Brazil president's eldest son diagnosed with coronavirus |
| 联合国拒绝美要求恢复制裁伊朗 | U.N. rejects U.S. demand to reinstate Iran sanctions |
| 赵立坚回应美防长说准备对抗中国 | Zhao Lijian responds to U.S. Defense Secretary's saying he's ready to confront China |
| 美国国会将调查蓬佩奥 | U.S. Congress to investigate Pompeo |
| 默克尔称欧盟期待与中方进一步合作 | Merkel says EU looks forward to further cooperation with China |
| 德国柏林再现反防疫措施示威 | Anti-epidemic measures demonstration in Berlin, Germany |
| 3 分钟混剪抗战史 | 3-Minute mix of war of resistance history |
| 特朗普公开攻击美军高层 | Trump publicly attacks top US military |
| 澳情报部门突击搜查中国驻澳记者住所 | Australian intelligence raided the residence of Chinese journalists in Australia |
| 94 岁英国女王将复工 | Queen to return to work at age 94 |
| 水门事件调查记者揭露特朗普淡化疫情 | Watergate investigative reporter exposes Trump's downplaying of epidemic |
| 中国关于联合国成立 75 周年立场文件 | China's position document on 75th anniversary of UN |
| 外交部回应美正秘密发展新核武器系统 | Foreign Ministry responds to US is secretly developing new nuclear weapons system |
| 王毅称中国从不干涉美国内政 | Wang Yi says China never interferes in US internal affairs |
| 廖国勋当选天津市市长 | Liao Guoxun elected mayor of Tianjin |
| 中国驻英大使接受 BBC 专访 | China's ambassador to Britain interviewed by BBC |
| 美国防长称希望年内访问中国 | U.S. Defense Secretary says he hopes to visit China within the year |

| | |
|---|---|
| 我国资源三号 03 星发射成功 | China's Resource III 03 star launched successfully |
| 北斗三号系统最后一颗组网卫星入网 | The last satellite of the Beidou-3 system is on the network |
| **福奇称家人遭受死亡威胁** | **Fauci says family under death threats** |
| **刘鹤与美贸易代表通话** | **Liu He speaks with US trade representative** |
| **外交部批澳方无理搜查 4 名驻澳中国记者** | **Foreign Ministry criticizes Australia's unreasonable search of four Chinese journalists in Australia** |
| **特朗普对气变加剧山火表示怀疑** | **Trump expresses doubts about gas change exacerbating mountain fires** |
| **特朗普称发推后经常感到后悔** | **Trump says he often regrets after tweeting** |
| **美国总统国家安全事务助理感染新冠** | **U.S. Assistant to the President for National Security Affairs infected with coronavirus** |
| 因疫情滞留新增食宿费用游客承担 | Stranded due to the epidemic new accommodation and food costs borne by tourists |
| 科技部要求严查违背科研诚信行为 | Ministry of Science and Technology requires strict investigation of violations of scientific research integrity |
| **外交部回应蓬佩奥对李登辉表示哀悼** | **Foreign Ministry responds to Pompeo's condolences to Lee Teng-hui** |
| **外交部回应驻日美军成防疫体系漏洞** | **Foreign Ministry responds to the U.S. military in Japan has become a loophole in the epidemic prevention system** |
| **印度再禁 15 款中国 APP** | **India bans 15 more Chinese APPs** |
| 人社部支持发展特色小店 | The Ministry of Human Resources and Social Security supports the development of special small stores |
| 张玉环回应申请国家赔偿 | Zhang Yuhuan responds to application for state compensation |
| **崔天凯回应多个中美热点话题** | **Cui Tiankai responded to a number of China-US hot topics** |
| **俄罗斯副总理新冠阳性** | **Russian deputy prime minister's coronavirus positive** |
| **拜登首度携竞选搭档亮相** | **Biden's first appearance with his running mate** |
| 新时代交通强国铁路先行规划纲要 | The new era of strong transportation railroad first planning outline |
| 杨克勤受贿案一审开庭 | Yang Keqin's bribery case in the first trial |
| **近 7 成受访美企对中国市场前景乐观** | **Nearly 70% of surveyed U.S. companies optimistic about China market outlook** |
| **葡萄牙总统跳海救人** | **Portuguese president jumped into the sea to save people** |
| **安倍晋三进入庆应大学医院** | **Shinzo Abe enters Keio University Hospital** |
| **外交部回应特朗普再签涉 TikTok 行政令** | **Foreign Ministry responds to Trump signing another executive order involving TikTok** |
| **特朗普反击奥巴马夫人** | **Trump hits back at Mrs. Obama** |

| | |
|---|---|
| 香港将向世贸组织申诉 | Hong Kong will appeal to the WTO |
| 拜登成为美国民主党总统候选人 | Biden becomes U.S. Democratic Party presidential candidate |
| 英法德称美无权重启联合国对伊制裁 | Britain, France and Germany say US has no right to reopen UN sanctions against Iran |
| 外交部回应 TikTok 起诉美国政府 | Foreign Ministry responds to TikTok's lawsuit against U.S. government |
| 香港反对派议员林卓廷及许智峰被捕 | Hong Kong opposition lawmakers Lam Cheuk-ting and Hui Chi-fung arrested |
| 美国威斯康星州进入紧急状态 | U.S. Wisconsin enters state of emergency |
| 美国制裁研制新冠疫苗的俄罗斯机构 | U.S. sanctions Russian agency that developed coronavirus vaccine |
| 彭斯正式接受共和党副总统候选人提名 | Pence formally accepts Republican nomination for vice president |
| 教育部将加强跟踪指导校园疫情防控 | Education Ministry to strengthen follow-up on campus outbreak prevention and control |
| 国防部回应美拟在日部署中导 | Defense Ministry responds to proposed US deployment of intermediate range missiles in Japan |
| 毛里求斯爆发千人游行 | Thousands march in Mauritius |
| 菅义伟决定竞选自民党总裁 | Yoshihide Suga decides to run for LDP president |
| 需要人陪 | The need for company |
| 特朗普连发 4 推谴责波特兰示威 | Trump condemns Portland protest in 4 tweets |
| 教育部发布高校命名最新规范 | Ministry of Education releases latest norms for naming colleges and universities |
| 美国称不加入与世卫有关的疫苗开发 | U.S. says it won't join WHO-related vaccine development |
| 特朗普承认曾淡化新冠疫情严重性 | Trump admits to having downplayed severity of coronavirus outbreak |
| 中方要求美国不要干涉中国内政 | China asks U.S. not to interfere in China's internal affairs |
| 菅义伟当选日本自民党总裁 | Yoshihide Suga elected president of Japan's Liberal Democratic Party |
| 英国暂停与香港间的引渡条约 | UK suspends extradition treaty with Hong Kong |
| 美国驻成都总领事馆现场情况 | U.S. Consulate General in Chengdu on site |
| 王毅称中国将坚定而理性地回应美国 | Wang Yi says China will respond firmly and rationally to U.S. |
| 中方坚决反对人为制造所谓新冷战 | China firmly opposes artificially created so-called new cold war |
| 全国危化品储存安全专项检查整治 | National hazardous chemical storage safety special inspection and rectification |
| 外交部回应中加关系遭遇困难 | Foreign Ministry responds to difficulties in China-Canada relations |
| 普京宣布俄首个新冠疫苗注册 | Putin announces registration of Russia's first coronavirus vaccine |
| 特朗普遭 52 家科技巨头联名起诉 | Trump was jointly sued by 52 tech giants |

| | |
|---|---|
| 检方通报李心草溺亡案件 | The prosecution informed Li Xincao drowning case |
| **多家美国企业联手抵制特朗普封杀微信** | **Many U.S. companies join forces to resist Trump's blocking of WeChat** |
| **中国驻休斯敦总领事馆全体馆员归国** | **Chinese Consulate General in Houston returns all members to China** |
| **外交部回应美方宣布取消中美经贸谈判** | **Foreign Ministry responds to U.S. announcement to cancel U.S.-China economic and trade talks** |
| **特朗普呼吁抵制固特异轮胎** | **Trump calls for boycott of Goodyear tires** |
| **拜登称如果当选将要求全美戴口罩** | **Biden says he will ask all Americans to wear masks if elected** |
| **特朗普 83 岁姐姐录音** | **Trump's 83-year-old sister recording** |
| **特朗普被正式提名为共和党总统候选人** | **Trump was officially nominated as the Republican presidential candidate** |
| **国防部回应美军机擅闯我演习禁飞区** | **The Ministry of Defense responded to the U.S. military aircraft trespassed exercise no-fly zone** |
| **美宣布制裁 24 家参与南海建岛中企** | **The United States announced sanctions against 24 Chinese companies involved in the construction of islands in the South China Sea** |
| **安倍称辞职如断肠之痛** | **Abe says resignation like a broken heart** |
| **外交部回应向南海发射导弹报道** | **Foreign Ministry responds to reports of missile launches into South China Sea** |
| **日本政府计划 9 月 17 日选出新首相** | **Japanese government plans to elect new prime minister on Sept. 17** |
| **外交部副部长约见捷克驻华大使** | **Vice Foreign Minister meets with Czech Ambassador to China** |
| **王毅点名警告捷克参议长你过线了** | **Wang Yi warns Czech Senate President by name crossed the line** |
| **印度宣布再禁用 118 款中国 App** | **India announces ban on 118 more Chinese apps** |
| **印巴交火一名印度军官死亡** | **India, Pakistan exchange fire, one Indian officer dead** |
| **英国拟向欧盟发通牒** | **UK to issue ultimatum to EU** |
| **外交部驳斥美称中国操纵湄公河水资源** | **Foreign Ministry refutes US claim that China manipulates Mekong water resources** |
| **阿富汗副总统遇袭受轻伤** | **Afghan vice president slightly wounded in attack** |
| **中俄发布外交部长联合声明** | **China, Russia issue joint statement by foreign ministers** |
| **蓬佩奥称越来越多国家支持美方世界观** | **Pompeo says more countries support US worldview** |
| **日本新首相菅义伟首次记者会** | **Japan's New Prime Minister Yoshihide Suga's First Press Conference** |
| 外交部新任发言人汪文斌 | New Foreign Ministry spokesman Wang Wenbin |
| 中组部划拨 1.2 亿元党费用于防汛救灾 | Ministry allocates 120 million yuan in party expenses for flood relief |

| | |
|---|---|
| **外交部回应英国将暂停与香港引渡条约** | **Foreign Ministry responds to UK to suspend extradition treaty with Hong Kong** |
| 坚决防止家长他人代劳等参赛造假行为 | Resolutely prevent falsification of participation by parents and others on their behalf |
| **中方将考虑不承认 BNO 为有效旅行证件** | **China will consider not recognizing BNO as a valid travel document** |
| 文化和旅游部副部长李金早被查 | Vice Minister of Culture and Tourism Li Jinzao investigated |
| **芬兰女总理与相恋 16 年男友结婚** | **Finnish female PM marries boyfriend of 16 years in love** |
| 刘永坦捐出最高科技奖 800 万奖金 | Liu Yongtan donates 8 million prize money top science and technology award |
| **上半年 5816 人放弃美国国籍** | **5,816 people renounce U.S. citizenship in first half year** |
| **外交部说中国人民是吓不倒的** | **Foreign Ministry says Chinese people can't be intimidated** |
| 十四五规划编制工作开展网上意见征求 | The 14th Five-Year Plan preparation work to carry out online opinion solicitation |
| **王毅赴机场迎接驻休斯敦总领事馆馆员** | **Wang Yi Goes to Airport to Greet Consulate General in Houston** |
| **美国没资格要求安理会恢复对伊制裁** | **U.S. not qualified to ask Security Council to restore sanctions against Iran** |
| 中国人民警察警旗式样 | Chinese people's police flag style |
| **越南逮捕 21 名中国通缉犯** | **Vietnam arrests 21 wanted Chinese criminals** |
| **东部战区台海演练针对的就是台独** | **Eastern Zone of War's Taiwan Strait drills target Taiwan independence** |
| **安倍晋三就辞职举行记者会** | **Shinzo Abe holds press conference on resignation** |
| 中央第七次西藏工作座谈会 | The seventh central Tibetan work symposium |
| **特朗普说自己的签名值 1 万美元** | **Trump says his signature is worth $10,000** |
| **特朗普考虑限制中国留学生** | **Trump considers restricting Chinese students** |
| 袁家军 | Yuan Jiajun |
| **中印国防部长在莫斯科举行会晤** | **China, India defense ministers meet in Moscow** |
| **特朗普政府考虑将中芯国际列入黑名单** | **Trump Administration Considers Blacklisting SMIC** |
| **肯塔基州两支游行队伍现场对峙** | **Two Kentucky marches face off on site** |
| **美驻华大使布兰斯塔德将离任** | **U.S. Ambassador to China Branstad to resign** |
| 王宁当选福建省省长 | Wang Ning elected governor of Fujian province |
| **台湾新竹空军基地内直升机坠落** | **Taiwan's Hsinchu Air Force base helicopter crash** |
| **特朗普自夸疫情发布会收视率超高** | **Trump's boast of epidemic conference gets superb ratings** |
| **外交部披露美私拆中方外交邮袋细节** | **Foreign Ministry discloses details of U.S. private demolition of Chinese diplomatic pouch** |

| | |
|---|---|
| 中方回应美方强行进入驻休斯敦总领馆 | **China responds to U.S. forcible entry into Consulate General in Houston** |
| 印度将审核 275 款中国 APP | **India to check 275 Chinese apps** |
| 杨洁篪谈中美关系 | **Yang Jiechi on China-US Relations** |
| 国旗法国徽法迎来重要修改 | National Flag Law and National Emblem Law See Important Amendments |
| 香港特区第六届立法会继续履行职责 | **HongKongSAR's 6th Legislative Council continues to perform its duties** |
| 美国考虑禁止疑似感染新冠公民回国 | **U.S. considers banning citizens suspected of infecting coronavirus from returning home** |
| 特朗普称哈里斯和拜登都是社会主义者 | **Trump says Harris and Biden are both socialists** |
| 特朗普考虑大选后撤换国防部长 | **Trump considers replacing defense secretary after election** |
| 白俄罗斯 | **Belarus** |
| 特朗普再次反对邮寄选票 | **Trump again opposes mail-in ballots** |
| 第三次国家药品集中采购 | The third national drug centralized procurement |
| 赵立坚双引号手势回应澳反华机构谣言 | **Zhao Lijian's double-quote gesture in response to rumors of anti-Chinese institutions in Australia** |
| 广州市原常务副市长苏泽群被查 | Former executive vice mayor of Guangzhou Su Zequn investigated |
| 俄外长称不会拒绝与中国开展 5G 合作 | **Russian foreign minister says won't reject 5G cooperation with China** |
| 安倍经济学 8 年回顾 | **Abe economics 8-year review** |
| 加拿大抗议者推倒首任总理雕像 | **Canadian protesters push down statue of first PM** |
| 在线旅游网站不得大数据杀熟 | Online travel sites must not big data-enabled price discrimination against existing customers |
| 外交部回应印度士兵在边界冲突中丧生 | **Foreign Ministry responds to Indian soldier killed in border clash** |
| 美方官员视中国留学生为间谍是妄想症 | **US officials see Chinese students as spies is paranoid** |
| 被绑架的 17 头牦牛 | 17 yaks kidnapped |
| 对美加征关税商品第一次排除延期清单 | **First exclusion extension list of goods subject to tariff increase against US** |
| 外交部回应菅义伟当选日本首相 | **Foreign Ministry responds to Yoshihide Suga's election as Japanese PM** |



\end{document}